# Decay and renormalization of a longitudinal mode in a quasi-two-dimensional antiferromagnet[‡]

Seung-Hwan Do,[1,†,*] Hao Zhang,[1,2,†] Travis J. Williams,[3] Tao Hong,[3] V. Ovidiu Garlea,[3] J. A. Rodriguez-Rivera,[4,5] Tae-Hwan Jang,[6] Sang-Wook Cheong,[6,7] Jae-Hoon Park,[6,8] Cristian D. Batista,[2] and Andrew D. Christianson[1]

[1]Materials Science and Technology Division, Oak Ridge National Laboratory, Oak Ridge, Tennessee 37831, USA

[2]Department of Physics and Astronomy, University of Tennessee, Knoxville, Tennessee 37996, USA

[3]Neutron Scattering Division, Oak Ridge National Laboratory, Oak Ridge, Tennessee 37831, USA

[4]Department of Materials Sciences, University of Maryland, College Park, Maryland 20742, USA

[5]NIST Center for Neutron Research, Gaithersburg, Maryland 20899, USA

[6]MPPHC-CPM, Max Planck POSTECH/Korea Research Initiative, Pohang 37673, Republic of Korea

[7]Rutgers Center for Emergent Materials and Department of Physics and Astronomy, Rutgers University, Piscataway, New Jersey 08854, USA

[8]Department of Physics, Pohang University of Science and Technology, 37673, Republic of Korea

†These authors contributed equally.

*To whom correspondence should be addressed. E-mail: doh1@ornl.gov

## Abstract

An ongoing challenge in the study of quantum materials, is to reveal and explain collective quantum effects in spin systems where interactions between different modes types are important. Here we approach this problem through a combined experimental and theoretical study of interacting transverse and longitudinal modes in an easy-plane quantum magnet near a continuous quantum phase transition. Our inelastic neutron scattering measurements of $Ba_2FeSi_2O_7$ reveal the emergence, decay, and renormalization of a longitudinal mode throughout the Brillouin zone. The decay of the longitudinal mode is particularly pronounced at the zone center. To account for the many-body effects of the interacting low-energy modes in anisotropic magnets, we generalize the standard spin-wave theory. The measured mode decay and renormalization is reproduced by including all one-loop corrections. The theoretical framework developed here is broadly applicable to quantum magnets with more than one type of low energy mode.

[‡] This manuscript has been authored by UT-Battelle, LLC under Contract No. DE-AC05-00OR22725 with the U.S. Department of Energy. The United States Government retains and the publisher, by accepting the article for publication, acknowledges that the United States Government retains a non-exclusive, paid-up, irrevocable, worldwide license to publish or reproduce the published form of this manuscript, or allow others to do so, for United States Government purposes. The Department of Energy will provide public access to these results of federally sponsored research in accordance with the DOE Public Access Plan (http://energy.gov/downloads/doe-public-access-plan).

**Introduction**

One of the strongest signatures of collective quantum behavior is the spontaneous quasiparticle decay in interacting bosonic systems, as observed in superfluids[1-3] and quantum magnets[4-8]. In the latter case, spontaneous magnon decay has been studied in a growing number of lattice geometries and model systems where large quantum fluctuations enhance this many body effect[9,10]. A key finding of these studies is that the strong decay process is accompanied by a significant renormalization of the overall spectrum[11-16]. This spectral renormalization leads to measurable effects in the thermal dynamic and transport properties[17], which are inexplicable without considering the renormalization of the quasiparticle mass. At the same time, the renormalization of the spectra opens an avenue to understand quantum systems since the renormalized single-magnon dispersion provides a stringent test for theories that attempt to describe the magnon decay. In other words, approaches that do not fully incorporate these many-body effects will not yield correct values of the interaction parameters extracted from experimental studies.

An important question is how to understand quasiparticle decay in quantum magnets when there is more than one type of low energy mode, *i.e.* when the parent particles are not of the same type as the daughter particles. Anisotropic magnets with spin $S \geq 1$ provide a common example of this situation. The additional fluctuations (quadrupolar for $S \geq 1$, octupolar for $S \geq 3/2$, etc.) can generate modes which are not captured by standard SU(2) approaches at the linear level. Rather, the physics is more conveniently described in terms of generalized SU($N$) spin wave theory, where the low energy modes are described by $N-1$ distinct bosons[18]. For example, anisotropic $S = 1$ systems where both transverse and longitudinal modes are expected, have been previously treated by linear SU(3) theories[17-23]. While linear SU($N$) approaches capture the correct number of low energy modes, they are unable to reproduce the quasi-particle decay and renormalization generated by the interaction between these modes. To capture these effects requires going beyond the linear level and thus an objective of this paper is to generalize the $1/S$-expansion of the SU(2) treatment to SU(3) in order to account for the quasi-particle decay and renormalization produced by the interaction (non-linear) terms using the quintessential example of interacting longitudinal and transverse modes for an $S = 1$ easy-plane quantum magnet as a test case.

In easy-plane quantum magnets, phase transitions can be driven by either fluctuations of the phase or the amplitude of the order parameter[24]. The phase fluctuations are the transverse modes of the order parameter (Goldstone modes in the long wavelength limit), whereas amplitude fluctuations correspond to the longitudinal modes. Due to the gapless nature of the Goldstone transverse modes, the longitudinal or "Higgs" mode is kinematically allowed to decay into two transverse modes. This decay becomes more significant in low-dimensional systems. Indeed, the longitudinal mode in two-dimensional (2D) antiferromagnets was originally assumed to be overdamped due to an infrared divergence of the imaginary part of the longitudinal susceptibility[25,26]. However, more recent theoretical work predicted that the longitudinal peak should remain visible even in 2D[27-33]. One aspect of this problem, which has not been emphasized in previous works, is that the rather strong decay of the longitudinal mode is accompanied by a significant renormalization of the gap and the dispersion of the modes. As noted above, this additional many-body effect provides a hard test for theories that attempt to reproduce the measured decay of the Higgs mode.

As a starting point to understand the physics described above, we focus on the quasi-2D Heisenberg square lattice with effective $S = 1$ with an antiferromagnetic exchange coupling ($\tilde{J}$) and a strong easy-plane single-ion anisotropy ($\tilde{D}$). In this case, $\alpha = \tilde{J}/\tilde{D}$ can be viewed as a tuning parameter that can be used to drive a system from a quantum paramagnet (QPM) to an antiferromagnet (AFM) with an intervening QCP as shown in the Fig. 1. Near the QCP, spontaneous symmetry breaking produces two transverse modes (one of them is a Goldstone mode) and a longitudinal Higgs mode. The longitudinal mode is unstable with respect to decay into a pair of transverse modes resulting in an intrinsic line-broadening[9,34].

In this paper, we use inelastic neutron scattering to study the spin excitation spectrum of $Ba_2FeSiO_7$. The high-quality neutron scattering data reveals a complex spectrum where transverse modes are resolution limited, whereas a longitudinal mode displays significant **Q**-dependent broadening throughout the Brillouin zone, demonstrating the importance of quasiparticle decay even away from the long wavelength limit. The neutron scattering results further show that the longitudinal mode has a very small gap clearly demonstrating that $Ba_2FeSiO_7$ is relatively close to a QCP. To understand the inelastic neutron scattering data, we implement a generalized SU(3) spin wave calculation[17,18,22] and compute the low-energy excitation spectrum of an effective low-energy spin $S = 1$ model. After demonstrating that the generalization of the well-known $1/S$-expansion of the SU(2) spin wave theory[35-41] is simply a loop expansion[42] of the SU(3) spin wave theory, we show that the one-loop correction is enough to account for the broadening of the longitudinal mode and the large renormalization of the gap and the dispersion of this mode. We further show that not including the one-loop corrections results in Hamiltonian parameters that place the exact ground state of the spin Hamiltonian for $Ba_2FeSi_2O_7$ on the nonmagnetic side of the QCP–contrary to experimental fact. This provides a dramatic demonstration of the importance of including renormalization effects, where the linear spin-wave calculation overestimates the stability range of the magnetically ordered state. The fact that the one-loop correction can simultaneously account for the real and imaginary part of the self-energy of the longitudinal mode, as well as of the renormalization the transverse mode dispersion, confirms that the easy plane quantum magnet $Ba_2FeSi_2O_7$ is an ideal platform for studying many-body effects in the proximity of the O(2) QCP.

**Results**

**Model Material**

Figure 2a illustrates the crystal structure of $Ba_2FeSi_2O_7$ comprising layers of $FeSi_2O_7$ separated by Ba atoms. As shown in Fig. 2b, the $FeO_4$ tetrahedra of the $FeSi_2O_7$ layer are connected via $SiO_4$ polyhedra and the two adjacent $Fe^{2+}$ atoms are coupled through the super-exchange interaction, $J$, that is mediated by the two oxygen ligands (red dashed line in Fig. 1b). The resulting square lattice of magnetic moments are vertically stacked along the $c$-axis, leading to a quasi-2D simple tetragonal spin-lattice.

A detailed description of the single-ion state of the $Fe^{2+}$ ion is given in Note 1 of the Supplementary Information. The combination of a relatively large spin-orbit coupling ($\lambda \sim 20$ meV) and a dominant tetrahedral crystal field ($\Delta_{Td}$), splits the free-ion levels, $^5D$ ($L = 2, S = 2$), into several multiplets. The lowest energy $S = 2$ multiplet has a significant orbital character due to the finite spin-orbit coupling, that combined with the tetragonal distortion ($\delta_{Tetra}$) by large compression of the FeO$_4$ tetrahedra leads to a rather strong easy-plane single-ion anisotropy[43,44]. The five $S = 2$ energy levels are then split into a singlet $S^z = 0$ ground state and two excited $S^z = \pm 1$ and $S^z = \pm 2$ doublets with energies $D$ and $4D$, respectively (see Fig. S1a of the Supplementary Information). Because the gap $D$ of the $S^z = \pm 1$ doublet is four times smaller than the gap of the $S^z = \pm 2$ doublet and the dominant super-exchange interaction $J$ is smaller than $D/4$ in Ba$_2$FeSi$_2$O$_7$, the low-energy spectrum is well captured by projecting the $S = 2$ spin Hamiltonian into the $S^z = 0$ and $S^z = \pm 1$ low-energy states.

The resultant $S = 1$ effective spin Hamiltonian describes the competition between a QPM ($\tilde{J} \ll \tilde{D}$) with each spin of the lattice having dominant $S^z = 0$ character, and a collinear AFM state ($\alpha = \tilde{J}/\tilde{D} > \alpha_c$) with staggered magnetization in the $ab$-plane (see Fig. 2b). Ba$_2$FeSi$_2$O$_7$ turns out to be on the antiferromagnetic side with a Néel temperature $T_N$=5.2 K[44]. Below $T_N$, the spins order antiferromagnetically with propagation vector $\mathbf{Q_m}$=(1,0,0.5), corresponding to $(\pi, \pi, \pi)$ as shown in Fig. 2c. The magnetic moments are highly confined in the $ab$-plane due to easy-plane anisotropy, giving rise to the magnetic structure shown in Fig. 2b. A neutron diffraction study on a powder sample revealed a significantly reduced ordered moment of 2.95 $\mu_B$, which is only 63% of the full moment of 4.36 $\mu_B$ ($g_{ab} = 2.18$) expected for an $S = 2$ spin[44], suggesting the proximity to the quantum critical point. Additionally, as described in further detail below, our analysis confirms that $\alpha = \tilde{J}/\tilde{D} \sim 0.184$ is close to the critical value, $\alpha_c^{2D}$=0.18 and $\alpha_c^{3D}$=0.1 for 2D and 3D respectively, obtained from quantum Monte Carlo simulations[22].

The spin excitations of Ba$_2$FeSi$_2$O$_7$ are generically described by an antiferromagnetic $S = 2$ spin Hamiltonian on a simple tetragonal lattice:

$$\mathcal{H} = J \sum_{\langle \mathbf{r},\mathbf{r'} \rangle} \left[ S_\mathbf{r}^x S_\mathbf{r'}^x + S_\mathbf{r}^y S_\mathbf{r'}^y + \Delta S_\mathbf{r}^z S_\mathbf{r'}^z \right]$$
$$+ J' \sum_{\langle\langle \mathbf{r},\mathbf{r'} \rangle\rangle} \left[ S_\mathbf{r}^x S_\mathbf{r'}^x + S_\mathbf{r}^y S_\mathbf{r'}^y + \Delta' S_\mathbf{r}^z S_\mathbf{r'}^z \right] \quad (1)$$
$$+ D \sum_\mathbf{r} (S_\mathbf{r}^z)^2.$$

The bracket $\langle \mathbf{r},\mathbf{r'} \rangle$ ($\langle\langle \mathbf{r},\mathbf{r'} \rangle\rangle$) indicates that the sum runs over intralayer (interlayer) nearest-neighbor spins with isotropic super-exchange interaction $J(J')$. $\Delta(\Delta')$ is the intralayer (interlayer) uniaxial anisotropy and the last term represents the easy-plane single-ion anisotropy ($D > 0$).

In the large $D/J$ limit, the $S^z = \pm 2$ doublet is separated from the $S^z = \pm 1$ doublet by an energy gap $3D$. The low-energy subspace of magnetic excitations can then be further reduced by projecting out the $S^z = \pm 2$ doublet. The reduced low-energy Hamiltonian $\mathcal{H}_{eff}$ results from projecting $\mathcal{H}$ onto the low-energy subspace $S$ spanned by the triplet of states with $S^z = 0, \pm 1$: $\mathcal{H}_{eff} = \mathcal{P}_S \mathcal{H} \mathcal{P}_S$. The resulting effective spin $S = 1$ Hamiltonian is

$$\mathcal{H}_{eff} = \tilde{J} \sum_{\langle \mathbf{r},\mathbf{r}'\rangle} [s_{\mathbf{r}}^x s_{\mathbf{r}'}^x + s_{\mathbf{r}}^y s_{\mathbf{r}'}^y + \tilde{\Delta} s_{\mathbf{r}}^z s_{\mathbf{r}'}^z]$$
$$+ \tilde{J}' \sum_{\langle\langle \mathbf{r},\mathbf{r}'\rangle\rangle} [s_{\mathbf{r}}^x s_{\mathbf{r}'}^x + s_{\mathbf{r}}^y s_{\mathbf{r}'}^y + \tilde{\Delta}' s_{\mathbf{r}}^z s_{\mathbf{r}'}^z] \quad (2)$$
$$+ \tilde{D} \sum_{\mathbf{r}} (s_{\mathbf{r}}^z)^2.$$

with $\tilde{J} = 3J$, $\tilde{J}' = 3J'$, $\tilde{\Delta} = \Delta/3$, $\tilde{\Delta}' = \Delta'/3$ and $\tilde{D} = D$. As we will see below, this simple effective Hamiltonian can explain not only the in-plane antiferromagnetic ordering observed in $Ba_2FeSi_2O_7$ (see Fig. 2b), but also the spectra of quasi-particle excitations, including rather strong renormalization effects due to proximity to the QCP.

**Inelastic neutron scattering**

To investigate the spin excitation spectrum in $Ba_2FeSi_2O_7$, we performed inelastic neutron scattering using two instruments; the cold neutron triple-axis spectrometer (CTAX) at the High Flux Isotope Reactor, and the time-of-flight hybrid spectrometer (HYSPEC) at the Spallation Neutron Source at Oak Ridge National Laboratory[45]. An overview of the inelastic neutron scattering results is presented in Fig. 3 through contour maps of the neutron scattering intensity, $I(\mathbf{Q}, \omega)$, along $[H, H, 0.5]$ and $[H, 0, 0.5]$. For both spectra, strongly dispersive spin excitations extending up to energy~2.7 meV are observed. Whereas the dispersion along $[0, 0, L]$-direction is weak with a bandwidth of ~0.5 meV (see Note 4 in the Supplementary Information), which is expected for spin excitations of a quasi-two-dimensional spin system.

There are several distinct features in the inelastic neutron scattering data. An intense spin wave excitation emanates from the magnetic zone center (ZC), $\mathbf{Q}=(1, 0, 0.5)$, which arises due to the in-phase oscillation between $Fe^{2+}$ spins in the plane. We refer to this mode as $T_1$. Along the $[H, 0, 0.5]$ direction towards the zone boundary (ZB) at $\mathbf{Q}=(0, 0, 0.5)$ the $T_1$-mode reaches its maximum energy of ~2.5 meV. Another weak, but sharp mode, is visible along $[H, 0, 0.5]$ with an energy of 2.5 meV at the ZC. We refer to this mode as $T_2$. These two modes are expected for a strong easy-plane antiferromagnet, where transverse magnons split into gapless in-plane fluctuations ($T_1$-mode) and gapped out-of-plane fluctuations ($T_2$-mode). The finite value of the energy gap of the out-of-plane fluctuation at the ZC is associated with the strength of the easy-plane single-ion anisotropy[46].

The $T_1$ and $T_2$ transverse magnon modes are also observed along the $[H, H, 0.5]$ direction in Fig. 3d. Noticeably, an additional sharp mode is observed at the top of the $T_1$-mode. This mode is visible along the entire Brillouin zone boundary. We refer to this additional mode as '$L$'-mode. The $L$-mode is visible in the spectra along $[H, 0, 0.5]$ as well, however, it exhibits dramatic line-broadening near ZC. To demonstrate more clearly the **Q**-dependence of the modes, Fig. 4 shows cuts at constant momentum transfers for multiple points along $[H, 0, 0.5]$ and $[H, H, 0.5]$. Two pronounced peaks, corresponding to the $T_1$- and $L$-modes, remain sharp along the ZB (Fig. 4b). As already noted, the situation is very different near the ZC where the $L$-mode is significantly broadened (Fig. 4a). We note that the $L$-mode remains a broad peak near the ZC, rather than a featureless excitation. To investigate the extent of the broadening effect, Gaussian line shapes for the $T_1$-, $T_2$-, and $L$-modes were fit to the individual cuts in Fig. 4. The line widths obtained from the fits are displayed in Fig. 7a-d. These results reveal that the $L$-mode is three times broader than the instrumental resolution at the ZC (see Fig. 4a), whereas it has comparable line width to instrumental resolution near the ZB.

**Generalized spin waves**

In this section we introduce a generalized SU(3) spin wave calculation[17,18,22,47], which is required to capture the two low-energy (longitudinal and transverse) modes of Ba$_2$FeSi$_2$O$_7$. Clearly, a linear treatment is not enough to capture the decay of the longitudinal mode into two transverse modes. Consequently, the main aim of this section is to lay the groundwork for introducing the loop expansion[42] (generalization of the $1/S$-expansion[35-41]) in the section describing the non-linear corrections.

To account for the transverse and longitudinal modes revealed by the INS experiment, the usual SU(2) spin-wave theory (SWT) must be generalized to SU(3)[18], by introducing the SU(3) Schwinger boson representation of the spin operators $S_\mathbf{r}^\nu = \mathbf{b}_\mathbf{r}^\dagger \mathcal{S}^\nu \mathbf{b}_\mathbf{r}$, where $\mathbf{b}_\mathbf{r} = (b_{\mathbf{r},+1}, b_{\mathbf{r},-1}, b_{\mathbf{r},0})^T$,

$$\mathcal{S}^x = \frac{1}{\sqrt{2}}(\lambda_4 + \lambda_6), \qquad \mathcal{S}^y = \frac{1}{\sqrt{2}}(\lambda_5 - \lambda_7), \qquad \mathcal{S}^z = \lambda_3, \qquad (3)$$

$\lambda_i$ are the Gel-Mann matrices and the Schwinger boson operators satisfy the local constraint

$$\sum_{m=\pm 1,0} b_{\mathbf{r},m}^\dagger b_{\mathbf{r},m} = M = 1. \qquad (4)$$

We note that the SU(3) Schwinger boson representation of the spin operators should not be confused with the Schwinger boson approximation[36,48-50], which is qualitatively different from the semi-classical approach that we describe below. The magnetically ordered state of Ba$_2$FeSi$_2$O$_7$ can be approximated by a product (mean-field) state of normalized SU(3) coherent states

$$|\psi_\mathbf{r}\rangle = \cos\theta|0\rangle + (\sin\theta\cos\phi|1\rangle + \sin\theta\sin\phi|-1\rangle)e^{i\mathbf{Q_m}\cdot\mathbf{r}}, \qquad (5)$$

where $\mathbf{Q_m} = (\pi,\pi,\pi)$ ((1, 0, 0.5) in the chemical lattice) is the AFM ordering wave vector. Although a general SU(3) coherent state is a parameterized by 4 independent parameters for degenerate representations[51], the two independent parameters $\theta$ and $\phi$ are enough to describe the collinear order under consideration. The three basis states $|m\rangle$ ($m = 0, \pm 1$) are represented by creating a boson with quantum number $m$ from the vacuum: $|m\rangle = b^\dagger_{\mathbf{r},m}|\varnothing\rangle$.

As in the usual spin wave theory, we introduce an SU(3) transformation that rotates the boson operators, $\widetilde{\boldsymbol{b}}_\mathbf{r} = U_\mathbf{r}\boldsymbol{b}_\mathbf{r}$, to a local basis that includes the coherent SU(3) state (5) as one of its three elements. This local transformation allows us to align the quantization axis with the direction of the local SU(3) order parameter. The spatial dependence of $U_\mathbf{r}$ can be removed by working in a twisted frame, where the original AFM order becomes a FM one. This can be done by rotating the spin reference frame of one of the two sublattices of the tetragonal lattice by an angle $\pi$ along the z-axis: $s^z_\mathbf{r} \to s^z_\mathbf{r}$, and $s^{x,y}_\mathbf{r} \to -s^{x,y}_\mathbf{r}$. In the new reference frame, the effective Hamiltonian (2) becomes

$$\widetilde{\mathcal{H}}_{eff} = \widetilde{J}\sum_{\langle\mathbf{r},\mathbf{r'}\rangle,\nu} a_\nu s^\nu_\mathbf{r} s^\nu_\mathbf{r'} + \widetilde{J}'\sum_{\langle\mathbf{r},\mathbf{r'}\rangle,\nu} b_\nu s^\nu_\mathbf{r} s^\nu_\mathbf{r'} + \widetilde{D}\sum_\mathbf{r}(s^z_\mathbf{r})^2, \qquad (6)$$

with $a_x = a_y = b_x = b_y = -1$, $a_z = \widetilde{\Delta}$ and $b_z = \widetilde{\Delta}'$, and the SU(3) transformation reads

$$U = \begin{pmatrix} -\sin\phi & \cos\phi & 0 \\ \cos\theta\cos\phi & \cos\theta\sin\phi & -\sin\theta \\ \sin\theta\cos\phi & \sin\theta\sin\phi & \cos\theta \end{pmatrix}. \qquad (7)$$

The bosonic representation of $\widetilde{\mathcal{H}}_{eff}$ is

$$\begin{aligned}\widetilde{\mathcal{H}}_{eff} = &\widetilde{J}\sum_{\langle\mathbf{r},\mathbf{r'}\rangle,\nu} a_\nu \widetilde{\boldsymbol{b}}^\dagger_\mathbf{r}\widetilde{\mathcal{S}}^\nu\widetilde{\boldsymbol{b}}_\mathbf{r}\widetilde{\boldsymbol{b}}^\dagger_{\mathbf{r'}}\widetilde{\mathcal{S}}^\nu\widetilde{\boldsymbol{b}}_{\mathbf{r'}} \\ &+ \widetilde{J}'\sum_{\langle\mathbf{r},\mathbf{r'}\rangle,\nu} b_\nu \widetilde{\boldsymbol{b}}^\dagger_\mathbf{r}\widetilde{\mathcal{S}}^\nu\widetilde{\boldsymbol{b}}_\mathbf{r}\widetilde{\boldsymbol{b}}^\dagger_{\mathbf{r'}}\widetilde{\mathcal{S}}^\nu\widetilde{\boldsymbol{b}}_{\mathbf{r'}} \\ &+ \widetilde{D}\sum_\mathbf{r}\left(1 - \widetilde{\boldsymbol{b}}^\dagger_\mathbf{r}\widetilde{\mathcal{A}}\widetilde{\boldsymbol{b}}_\mathbf{r}\right),\end{aligned} \qquad (8)$$

where $\widetilde{\mathcal{S}}^\nu = U\mathcal{S}^\nu U^\dagger$, $\widetilde{\mathcal{A}} = U\mathcal{A}U^\dagger$, and $\mathcal{A}_{\alpha\beta} = \delta_{\alpha,0}\delta_{\beta,0}$. Note that the unitary transformation (7) is chosen in such a way that the $\widetilde{b}_{\mathbf{r},0}$ boson is macroscopically occupied, namely $\langle\widetilde{b}_{\mathbf{r},0}\rangle = \langle\widetilde{b}^\dagger_{\mathbf{r},0}\rangle \simeq \sqrt{M}$. According to the constraint (4), $M = 1$ for the case of interest. However, we will keep using $M$ because $1/M$ is the parameter of the perturbative expansion that will be introduced below. Note that $M = 2S$ for the usual SU(2) spin wave theory. The main assumption behind the $1/M$ expansion is that $\langle\widetilde{b}^\dagger_{\mathbf{r},-1}\widetilde{b}_{\mathbf{r},-1}\rangle, \langle\widetilde{b}^\dagger_{\mathbf{r},+1}\widetilde{b}_{\mathbf{r},+1}\rangle \ll M$. Under this assumption, we can expand the spin operators $S^\mu$ and the quadrupolar operator $(S^z)^2$ in powers of $1/M$ (see Note 5 in the Supplementary Information). The resulting expansion of $\widetilde{\mathcal{H}}_{eff}$ is

$$\tilde{\mathcal{H}}_{eff} = M^2 \mathcal{H}^{(0)} + M H^{(2)} + M^{1/2} H^{(3)} + M^0 H^{(4)} + O(M^{-1}), \tag{9}$$

where the linear term $H^{(1)}$ vanishes because the parameters $\theta$ and $\phi$ in Eq. (5) are determined by minimizing the mean field energy

$$\begin{aligned}\mathcal{H}^{(0)} &= (2\tilde{J}\tilde{\Delta} + \tilde{J}'\tilde{\Delta}')\sin^4\theta\cos^2 2\phi \\ &\quad - \frac{1}{2}(2\tilde{J} + \tilde{J}')\sin^2 2\theta(1 + \sin 2\phi) + \tilde{D}\sin^2\theta.\end{aligned} \tag{10}$$

Since the AFM order is invariant under time reversal followed by one lattice translation, the states $|S_z = \pm 1\rangle$ must have equal weight in the mean field state (5), implying that $\phi = \pi/4$. By minimizing $\mathcal{H}^{(0)}$ with respect to $\theta$, we obtain

$$x \equiv \sin^2\theta = \frac{1}{2} - \frac{\tilde{D}}{8(2\tilde{J} + \tilde{J}')}. \tag{11}$$

The quadratic term $\mathcal{H}^{(2)}$ represents the generalized linear spin wave (GLSW) Hamiltonian. After Fourier transforming the bosonic operators,

$$\tilde{b}_{r\alpha} = \frac{1}{\sqrt{N_s}} \sum_{\mathbf{k}} \tilde{b}_{\mathbf{k}\alpha} e^{i\mathbf{k}\cdot\mathbf{r}}, \tag{12}$$

where $N_s$ is the number of sites, $\mathcal{H}^{(2)}$ can be brought into a compact form by introducing the Nambu spinor $\vec{b}_{\mathbf{k}} = (\tilde{b}_{\mathbf{k},+1}, \tilde{b}_{\mathbf{k},-1}, \tilde{b}^\dagger_{-\mathbf{k},+1}, \tilde{b}^\dagger_{-\mathbf{k},-1})^T$,

$$\mathcal{H}^{(2)} = \sum_{\mathbf{k}} \sum_{\alpha,\beta=\pm 1} \vec{b}^\dagger_{\mathbf{k}} \mathcal{H}^{(2)}(\mathbf{k}) \vec{b}_{\mathbf{k}}, \tag{13}$$

with

$$\mathcal{H}^{(2)}(\mathbf{k}) = \begin{pmatrix} \Delta_{\alpha\beta}(\mathbf{k}) & \Lambda_{\alpha\beta}(\mathbf{k}) \\ \Lambda_{\beta\alpha}(\mathbf{k}) & \Delta_{\beta\alpha}(\mathbf{k}) \end{pmatrix}. \tag{14}$$

The matrix elements are

$$\begin{aligned}\Delta_{\alpha\beta}(\mathbf{k}) &= \sum_{\nu}[(2a_\nu \tilde{J} + b_\nu \tilde{J}')(\tilde{S}^\nu_{\alpha\beta}\tilde{S}^\nu_{00} - (\tilde{S}^\nu_{00})^2 \delta_{\alpha\beta}) \\ &\quad + (\tilde{J}a_\nu \sum_{\nu'=x,y} \cos k_{\nu'} + \tilde{J}'b_\nu \cos k_z)\tilde{S}^\nu_{\alpha 0}\tilde{S}^\nu_{0\beta}] \\ &\quad - \frac{\tilde{D}}{2}(\tilde{\mathcal{A}}_{\alpha\beta} - \tilde{\mathcal{A}}_{00}\delta_{\alpha\beta}),\end{aligned} \tag{15}$$

$$\Lambda_{\alpha\beta}(\mathbf{k}) = \sum_{\nu} \tilde{S}^\nu_{\alpha 0}\tilde{S}^\nu_{\beta 0}[\tilde{J}a_\nu \sum_{\nu'=x,y} \cos k_{\nu'} + \tilde{J}'b_\nu \cos k_z]. \tag{16}$$

The collinear mean-field state (5) has a residual $Z_2$ symmetry associated with a $\pi$ rotation along the direction of the ordered moments (local $\tilde{z}$-axis). The bosonic operator $\tilde{b}^\dagger_{+1}$ picks up minus sign under this $Z_2$ symmetry because it creates the state with $\tilde{S}^z = -1$. In contrast, the bosonic operator $\tilde{b}^\dagger_{-1}$ remains invariant because it creates the state with $\tilde{S}^z = 0$. This symmetry analysis implies that the $\tilde{b}_{+1}$ and $\tilde{b}_{-1}$ bosons must be decoupled in $\mathcal{H}^{(2)}$ because a non-vanishing hybridization term would otherwise break this $Z_2$ symmetry:

$$\mathcal{H}^{(2)} = \sum_{\mathbf{k},\alpha=\pm 1} [A_{\mathbf{k},\alpha} \tilde{b}^\dagger_{\mathbf{k},\alpha} \tilde{b}_{\mathbf{k},\alpha} - \frac{B_{\mathbf{k},\alpha}}{2}(\tilde{b}_{-\mathbf{k},\alpha}\tilde{b}_{\mathbf{k},\alpha} + \tilde{b}^\dagger_{\mathbf{k},\alpha}\tilde{b}^\dagger_{-\mathbf{k},\alpha})] \quad (17)$$

with $\gamma^{xy}_{\mathbf{k}} = \cos(k_x) + \cos(k_y)$, $\gamma^{z}_{\mathbf{k}} = \cos(k_z)$ and the expressions for $A_{\mathbf{k},\alpha}$ and $B_{\mathbf{k},\alpha}$ are given in Note 5 of the Supplementary Information.

The diagonal form of $\mathcal{H}^{(2)}$,

$$\mathcal{H}^{(2)} = \sum_{\mathbf{k},\alpha=\pm 1} \omega_{\mathbf{k},\alpha}\left(\beta^\dagger_{\mathbf{k},\alpha}\beta_{\mathbf{k},\alpha} + \frac{1}{2}\right) \quad (18)$$

is then obtained by applying an independent Bogoliubov transformation for each bosonic flavor,

$$\tilde{b}_{\mathbf{k},\pm 1} = u_{\mathbf{k},\pm 1}\beta_{\mathbf{k},\pm 1} + v_{\mathbf{k},\pm 1}\beta^\dagger_{-\mathbf{k},\pm 1}, \quad (19)$$

with

$$u_{\mathbf{k},\pm 1} = \sqrt{\frac{1}{2}\left(\frac{|A_{\mathbf{k},\pm 1}|}{\omega_{\mathbf{k},\pm 1}} + 1\right)},$$

$$v_{\mathbf{k},\pm 1} = \frac{B_{\mathbf{k},\pm}}{|B_{\mathbf{k},\pm}|}\sqrt{\frac{1}{2}\left(\frac{|A_{\mathbf{k},\pm 1}|}{\omega_{\mathbf{k},\pm 1}} - 1\right)}. \quad (20)$$

The operators $\beta^\dagger_{\mathbf{k},\pm 1}$ create quasi-particles with energy

$$\omega_{\mathbf{k},\pm 1} = \sqrt{A^2_{\mathbf{k},\pm 1} - B^2_{\mathbf{k},\pm 1}}, \quad (21)$$

where $\omega_{\mathbf{k},+1}$ ($\omega_{\mathbf{k},-1}$) is the dispersion relation of the transverse (longitudinal) modes. The neutron scattering intensity $I(\mathbf{Q},\omega)$ is related to the spin-spin correlation function through

$$I(\mathbf{Q},\omega) \propto f^2(\mathbf{Q}) \sum_{\mu,\nu}\left(\delta_{\mu\nu} - \frac{\hat{Q}_\mu \hat{Q}_\nu}{Q^2}\right)$$
$$\times \frac{1}{2\pi N_s}\sum_{i,j}^{N_s} \int_{-\infty}^{+\infty} dt\, e^{i\omega t - i\mathbf{Q}\cdot(\mathbf{r}_i - \mathbf{r}_j)}\langle s^\mu_i(t) s^\nu_j(0)\rangle, \quad (22)$$

where $\mathbf{Q}$ is the momentum vector transfer, and $f(\mathbf{Q})$ is the magnetic form factor of $Fe^{2+}$. In the Discussion section, we will show that although the GLSW approach discussed in this section can reproduce the dispersion relations of all observed low-energy modes in $Ba_2FeSi_2O_7$, it cannot account for various interaction effects that are revealed by the INS experiments. To capture these effects, we must then include the next order terms in the $1/M$-expansion.

**Non-linear corrections**

In this section, we demonstrate that the generalization of the $1/S$-expansion is simply a loop expansion. Based on this result, we compute the one-loop corrections to the linear theory presented in the previous section. As we explain in the next section, the one-loop correction accounts for both the broadening and the energy renormalization of the longitudinal mode near the zone center.

After Fourier transforming and applying a Bogoliubov transformation, the cubic contributions to the generalized spin wave theory become

$$\mathcal{H}^{(3)} = \mathcal{H}_c^{(3)} + \mathcal{H}_l^{(3)}, \tag{23}$$

with

$$\begin{aligned}\mathcal{H}_c^{(3)} = \frac{1}{\sqrt{N_s}} \sum_{\mathbf{q}_i} \sum_{\alpha_i = \pm 1} \delta(\mathbf{q}_1 + \mathbf{q}_2 + \mathbf{q}_3) \\ \times [\frac{1}{3!} V_s^{(3)}(\mathbf{q}_{1,2,3}, \alpha_{1,2,3}) \beta_{\mathbf{q}_1, \alpha_1} \beta_{\mathbf{q}_2, \alpha_2} \beta_{\mathbf{q}_3, \alpha_3} \\ + \frac{1}{2!} V_d^{(3)}(\mathbf{q}_{1,2,3}, \alpha_{1,2,3}) \beta_{\bar{\mathbf{q}}_1, \alpha_1}^\dagger \beta_{\bar{\mathbf{q}}_2, \alpha_2}^\dagger \beta_{\mathbf{q}_3, \alpha_3} + h.c.],\end{aligned} \tag{24}$$

and

$$\begin{aligned}\mathcal{H}_l^{(3)} &= \frac{1}{\sqrt{N_s}} \sum_{\mathbf{q}} \sum_{\alpha = \pm 1} \left[V_l^{(3)}(\mathbf{q}, \mathbf{0}, \mathbf{q}; \alpha, -1, \alpha) \beta_{\mathbf{0}, -1}^\dagger + h.c.\right] \\ &= \sqrt{N_s} \sum_{\alpha = \pm 1} \left[V_{L, \alpha} \beta_{\mathbf{0}, -1}^\dagger + h.c.\right].\end{aligned} \tag{25}$$

Here $V_d^{(3)}$ and $V_s^{(3)}$ are the decay and sink vertices, respectively. The symmetry allowed cubic vertices are depicted in the second and third lines of Fig. 4. Note that, unlike the SU(2) case, collinear magnetic ordering does not preclude cubic terms in the expansion (9) of the generalized SU(N) spin wave theory with $N > 2$. For the SU(3) case under consideration, the residual $Z_2$ symmetry ($\pi$-rotation along the local $\tilde{z}$-axis) only requires that the $\tilde{b}_{+1}$ boson must appear an even number of times (e.g., $\tilde{b}_{+1}\tilde{b}_{+1}$ or $\tilde{b}_{+1}^\dagger \tilde{b}_{+1}^\dagger$) in the cubic terms. $\mathcal{H}_l^{(3)}$ is a linear term that originates from the normal-ordering of the cubic vertices. This term renormalizes the optimal value $\theta$ that was obtained from the minimization of $\mathcal{H}^{(0)}$. The integral of $V_l^{(3)}(\mathbf{q}; \alpha, -1)$ over the entire Brillouin zone is the so-called cubic-linear vertex, which is non-zero only for the longitudinal boson at the ordering wave vector $\mathbf{q}=0$ (in the twisted frame). The explicit forms of $V_{d,s}^{(3)}$ and $V_l^{(3)}$ are derived in Note 7 of the Supplementary Information.

We will now describe the construction of a systematic perturbative field theory that is controlled by $1/M$. This scheme can be applied to study anharmonicities starting from any generalized spin wave theory based on a Schwinger boson representation of the generators of SU($N$). The well-known $1/S$-expansion will be recovered for the particular case $N = 2$ and $M = 2S$. As we will demonstrate below, the $1/M$-expansion is just a particular example of the loop expansion that is commonly used to describe spontaneous symmetry breaking in particle theory[42]. The connection is more evident after noticing that $M$ becomes an overall prefactor of the rescaled Hamiltonian (Eq. (9)), $H = H_{eff}/M$, once we also rescale the bosonic fields according to $\bar{b}_{\mathbf{r},\nu} = \tilde{b}_{\mathbf{r},\nu}/\sqrt{M}$. Since the original interaction vertices $V^{(n)}$ ($n \geq 3$) scale as $V^{(n)} \sim (M)^{2-\frac{n}{2}}$, all vertices of the rescaled Hamiltonian $H(\{\bar{b}_{\mathbf{r},\nu}, \bar{b}^{\dagger}_{\mathbf{r},\nu}\})$ become of order $M$, while the propagator is still of order $1/M$. Thus, the order $p$ of a particular one-particle irreducible diagram is $V - I$, where $V$ is the number of vertices and $I$ is the number of internal lines (note that the frequency $\omega$ is of order $M^0$ because the quadratic contribution $\langle H^{(2)} \rangle$ is independent of $M$). Since the number of loops is $L = I - V + 1$ (Every vertex introduces a delta function that reduces the number of independent momenta by one, except for one delta function that is left over for overall energy momentum conservation), we obtain the desired result: $p = 1 - L$.

Let us rederive this result without rescaling the fields and the Hamiltonian. As we already mentioned, Eq. (9) tells us that the interaction vertices $V^{(n)}$ ($n \geq 3$) scale as $V^{(n)} \sim (M)^{2-\frac{n}{2}}$. The quasi-particle propagator

$$\mathcal{G}_{0,\alpha}(\mathbf{k}, i\omega) = (-i\omega + \omega_{\mathbf{k},\alpha})^{-1}, \alpha = \pm 1 \tag{26}$$

where $\omega$ is the Matsubara frequency, scales as $\mathcal{G}_{0,\alpha}(k) \sim M^{-1}$ because $\omega_{\mathbf{k},\alpha}$ is of order $M$ (see Eq. (9)). The dressed single-particle propagator is obtained from the Dyson equation,

$$\mathcal{G}^{-1}(\mathbf{k}, i\omega) = \mathcal{G}_0^{-1}(\mathbf{k}, i\omega) - \Sigma(\mathbf{k}, i\omega), \tag{27}$$

where $\Sigma(\mathbf{k}, i\omega)$ is the single-particle self-energy. At a given order in $M$, the dressed propagator includes two external legs, $L$ independent loops, $I$ internal lines (bare propagators $\mathcal{G}_0$) and $V_n$ interaction vertices of the type $V^{(n)}$. After a summation over the Matsubara frequency $\omega \sim M^1$, each loop gives a contribution of order $M^1$. Hence, the order $p$ of a particular one-particle irreducible diagram contributing to $\Sigma(\mathbf{k}, i\omega)$ is

$$p = L - I + \sum_{n\geq 3} V_n \left[2 - \frac{n}{2}\right]. \tag{28}$$

Since each internal line connects a pair of vertices, we have

$$\sum_{n\geq 3} nV_n = 2I + 2, \tag{29}$$

where $\sum_{n\geq 3} nV_n$ is the total number of lines. Furthermore, the number of loops is equal to the number of independent momentum integrals. From the conservation of momentum at each vertex, we have

$$L = I - \left[\sum_{n\geq 3} V_n - 1\right]. \tag{30}$$

By combining the above results, we obtain

$$p = 1 + \sum_{n\geq 3} V_n - \sum_{n\geq 3} \frac{nV_n}{2} = 1 - L, \tag{31}$$

implying that the order of a given diagram is determined by the number of loops.

The lowest-order $\mathcal{O}(M^0)$ Feynman diagrams are shown in Fig. 5. Since the inverse of the bare boson propagator is of order $\mathcal{O}(M^1)$, the remaining diagrams of order $\mathcal{O}(M^0)$ give a relative $1/M$-correction to the poles of the bare propagators. The real part of the new poles corresponds to the renormalized single-particle energy, whereas the imaginary part corresponds to the decay rate, which is responsible for the broadening of the quasi-particle peaks measured with INS.

The contributions to the self-energy from the decay and from the source diagrams shown in Fig. 6 are

$$\Sigma_\alpha^d(\mathbf{q}, i\omega) = \frac{1}{2N_s} \sum_{\mathbf{k},\alpha_1,\alpha_2=\pm 1} \frac{|V_d^{(3)}(\bar{\mathbf{k}}, \mathbf{k}+\bar{\mathbf{q}}, \mathbf{q}; \alpha_1, \alpha_2, \alpha)|^2}{i\omega - \omega_{\mathbf{k},\alpha_1} - \omega_{\mathbf{q}+\bar{\mathbf{k}},\alpha_2}}, \tag{32}$$

and

$$\Sigma_\alpha^s(\mathbf{q}, i\omega) = -\frac{1}{2N_s} \sum_{\mathbf{k},\alpha_1,\alpha_2=\pm 1} \frac{|V_s^{(3)}(\mathbf{k}, \bar{\mathbf{k}}+\bar{\mathbf{q}}, \mathbf{q}; \alpha_1, \alpha_2, \alpha)|^2}{i\omega + \omega_{\mathbf{k},\alpha_1} + \omega_{\mathbf{q}+\bar{\mathbf{k}},\alpha_2}}, \tag{33}$$

respectively.

Finally, the diagrams that appear in the last line for both panels of Fig. 6 arise from the normal ordering of the quartic term $\mathcal{H}^{(4)}$ in Eq. (9). These contributions simply renormalize the quadratic Hamiltonian:

$$\mathcal{H}_{NO}^{(4)} = \sum_{\mathbf{q},\alpha,\alpha'} [V_{\alpha\alpha'}^{(4,N)} \beta_{\mathbf{q},\alpha}^\dagger \beta_{\mathbf{q},\alpha'} + (V_{\alpha\alpha'}^{(4,A)} \beta_{-\mathbf{q},\alpha} \beta_{\mathbf{q},\alpha'} + h.c.)], \tag{34}$$

where $V_{\alpha\alpha'}^{(4,N)}$ ($V_{\alpha\alpha'}^{(4,A)}$) represents the normal (anomalous) contributions. Since $\mathcal{H}_{NO}^{(4)}$ is of order $M^0$, only the diagonal normal contribution arising from the normal vertex $V_{\alpha\alpha'}^{(4,N)} \delta_{\alpha,\alpha'}$ gives a relative correction of order $1/M$ to the bare single-particle energy given in Eq. (21) (the anomalous terms in Eq. (34) give a relative correction contribution order $1/M^2$). The derivation of $V_{\alpha\alpha}^{(4,N)}$ is included in Note 7 of the Supplementary Information.

We note the parallel between the decay, sink, and quartic diagrams that give the $1/M$-correction to the single-particle self-energy and the ones that appear in the $1/S$-expansion of the standard SU(2) spin wave theory of non-collinear Heisenberg magnets[11]. The main difference is that the SU(3) theory includes an extra bosonic flavor that enables more symmetry-allowed decay channels. In addition, the cubic-linear diagram exists even in absence of magnetic field because the magnitude of the ordered magnetic moment can be renormalized by changing the variational parameter $\theta$. These diagrams, shown in the third line of Fig. 6a and the fourth line of Fig. 6b, are obtained by contracting one of the legs of the decay vertex with the cubic-linear vertex shown in Fig. 5. By using the Feynman rules, the cubic-linear diagrams are calculated as

$$\Sigma_\alpha^{cl}(\mathbf{q}) = -\frac{1}{\omega_{\mathbf{0},-1}} ([V_d^{(3)}(\mathbf{0},\bar{\mathbf{q}},\mathbf{q};\alpha,-1,\alpha)]^* V_{L,\alpha} + h.c.). \tag{35}$$

By applying the analytic continuation $\omega \pm i\delta^+ \to i\omega$ and adopting the so-called on-shell approximation $\omega = \omega_\mathbf{q}$ for Eq. (32) and Eq. (33), the renormalized pole of the dressed propagator $\mathcal{G}$ is calculated as $\widetilde{\omega}_{\mathbf{q},\alpha} - i\widetilde{\Gamma}_{\mathbf{q},\alpha} = \omega_{\mathbf{q},\alpha} + V_{\mathbf{q},\alpha\alpha}^{(4,N)} + \Sigma_\alpha^{cl}(\mathbf{q}) + \Sigma_\alpha^s(\mathbf{q},\omega_{\mathbf{q},\alpha}) + \Sigma_\alpha^d(\mathbf{q},\omega_{\mathbf{q},\alpha})$, where the imaginary part of the pole $\widetilde{\Gamma}_{\mathbf{k},\alpha}$ arises from the decay term $\Sigma_\alpha^d$, that accounts for the observed broadening of the longitudinal mode in most regions of the BZ (see Fig. 3c and f) (the calculations are summarized in Note 9 of the Supplementary Information and Ref. 52). Moreover, the shift in the real part of the pole implies a corresponding renormalization in the model parameters. By fitting the neutron scattering data with the renormalized dispersion peaks $\widetilde{\omega}_{\mathbf{q},\alpha}$ at the ZC, we obtain the set of optimal Hamiltonian parameters listed as set $\mathcal{B}$ in Table 1 and discussed further below.

**Discussion**

**Comparison between experiment and theory**

To understand the spin excitation spectrum of $Ba_2FeSi_2O_7$ and demonstrate the importance of using the one-loop corrections, we start the comparison between experiment and theory with the GLSWT (i.e. without one-loop corrections). Figure 3b, and e show contour plots of $I(\mathbf{Q},\omega)$ (Eq. (22)) calculated with the GLSWT along the $[H, 0, 0.5]$-, and $[H, H, 0.5]$-direction, respectively. The Hamiltonian parameters (see set $\mathcal{A}$ in Table. 1) are extracted by fitting the measured positions of the quasi-particle peaks (Gaussian-fitted peak centers of the experimental data) at the ZC. The GLSWT reproduces the dispersion of the observed two transverse modes $T_1$ and $T_2$ along the $[H, 0, 0.5]$- and $[H, H, 0.5]$-directions (Fig. 3b and e). Noticeably, the calculated longitudinal mode closely reproduces the experimental dispersion of the '$L$'-mode, which demonstrates that the SU(3) spin wave theory describes the quasiparticles in $Ba_2FeSi_2O_7$.

Notably, the GLSWT does not reproduce the broadening and renormalization of the longitudinal modes observed in the inelastic neutron scattering data. This is because the effect arises from the decay of a longitudinal mode into two transverse modes that is induced by the cubic term $\mathcal{H}^{(3)}$ of the expansion (Eq. (9)). To capture this effect, the $1/M$-correction from the one-loop expansion (see Non-linear correction section) must be included. The GLSWT+one-loop correction can then describe the broadened spectrum of the longitudinal mode. The new Hamiltonian parameters, which are determined via the same procedure that is described above (see set $\mathcal{B}$ in Table 1), allow us to reproduce the observed spectrum (see Fig. 3c and f).

A more in-depth comparison between theory and experiment is shown in Figures 7a and b. These figures show the quasi-particle dispersions along the [$H$, 0, 0.5]-direction calculated with the GLSWT and GLSWT plus one-loop corrections compared to the measured dispersion. Near the ZC, $\mathbf{Q_m}$=(1, 0, 0.5) the energy of longitudinal mode obtained from the GLSWT is noticeably higher than the peak center of the measured '$L$'-mode (orange dots). The discrepancy in the dispersion is resolved by introducing the one-loop corrections. The real part of the self-energy renormalizes the energy of the longitudinal mode, leading to a better agreement with the observed peak positions near the ZC. At the same time, the imaginary part of the self-energy obtained from the decay diagrams, $\Sigma_\alpha^d$, leads to an intrinsic line-broadening of the longitudinal mode that is missing in the GLSWT. In Fig. 7b and d, the lower (upper) boundary of the red-shaded region is given by $\widetilde{\omega}_{\mathbf{k},-1}(\mp)\widetilde{\Gamma}_{\mathbf{k},-1}$, representing theoretical line-broadening of the longitudinal mode that is compared against the experimental FWHM (orange error bars). In particular, the above-mentioned effects are most striking at $\mathbf{Q_m}$=(1, 0, 0.5), therefore we present a comparison of the intensity line-cut at this momentum transfer in Fig. 7e. It is interesting to note that the energy shift of the transverse mode is also captured by the one-loop corrections.

After verifying that the one-loop corrections can simultaneously capture the broadening of the longitudinal mode and the energy shift of both the transverse and the longitudinal modes at the magnetic ZC, it is natural to ask if this also holds true far away from the ZC. Figure 7f, g are the intensity cuts for two representative points on the ZB. At a first glance, the peak centers of both modes are reasonably reproduced by the one-loop corrections. A more detailed analysis reveals that the experimental FWHM of both peaks is equal to the instrumental resolution. However, as illustrated in Fig. 8a, since the longitudinal modes are still inside the two-magnon continuum, the one-loop correction predicts an intrinsic broadening (black curves) in Fig. 7f, g.

To understand the origin of this discrepancy, we trace back the decay channel of the longitudinal mode on the zone boundaries. The two-magnon continuum at the zone edge starts at an energy equal to the sum of the single-magnon energies at the zone center and the zone boundary. Due to the U(1) symmetry of the effective Hamiltonian, the magnons are gapless at the zone center, implying that the onset of the two-magnon continuum coincides with the single magnon branch (see Fig. 8). In absence of U(1) symmetry, the magnon modes become gapped and the longitudinal mode does not need to lie inside the two-magnon continuum for arbitrary values of the wave vector (see Fig. 8b). A small magnon gap pushes the onset of the two-magnon continuum to be above the energy of the longitudinal mode at the zone boundaries. This modification of the two-magnon spectrum precludes the decay of the longitudinal mode near the zone boundary and explains the experimental observation. We then conjecture that the single-magnon dispersion is indeed gapped.

Unfortunately, it is difficult to extract the size of this gap from our INS data because of the large quasi-elastic scattering. Nevertheless, the analysis presented in Note 2 of the Supplementary

Information indicates that our data is indeed consistent with a finite spin gap. We note that the gap can be captured by working with the original spin $S = 2$ Hamiltonian (Eq. (1)). The tetragonal symmetry allows for a single-ion anisotropy term of the form $\mathcal{H}_A = A \sum_i [(S_i^x)^4 + (S_i^y)^4]$, which breaks the global U(1) symmetry, generating a finite gap for the transverse mode. However, when we project the original $S = 2$ Hamiltonian onto the low-energy space to obtain the effective spin $S = 1$ Hamiltonian (Eq. (2)), the term $\mathcal{H}_A$ simply renormalizes the single-ion anisotropy, implying that the low-energy model acquires an "emergent" U(1) symmetry that is absent in the original high-energy model. Lastly, we note that the energies of the longitudinal mode on the zone boundaries after the one-loop corrections are slightly lower than the measured values. This level of discrepancy can be attributed to the missing second order corrections $O\left(\frac{J^2}{3D}\right)$ to the low-energy model (2) or to missing terms in the original Hamiltonian (1). A simple analysis shows that a second nearest neighbor AFM interaction with $\tilde{J}_2 \sim 0.2\tilde{J}$ can account for this discrepancy. For simplicity, $\tilde{J}_2$ is not included in our calculation. Except for the discrepancy near the zone boundaries, the effective $S = 1$ model with one-loop corrections successfully captures most features of the INS data inside the BZ.

Finally, we emphasize that the loop expansion preserves the Goldstone mode that results from the spontaneous breaking of the emergent U(1) symmetry group of $\widetilde{\mathcal{H}}_{eff}$. More specifically, the $\mathcal{O}(M^0)$ correction to the real part of the self-energy vanishes for the Goldstone mode (see Note 8 in the Supplementary Information), although the individual contributions from the diagrams shown in Fig. 5 diverge as $1/q$ in the long-wavelength limit. We note that previous attempts of computing the decay of the longitudinal mode[34] have not accounted for the renormalization of the single-particle dispersion arising from the $1/M$-correction to the real part of the self-energy. This correction leads to a significant change in the extracted ratio $\alpha = \tilde{J}/\tilde{D}$ of Ba$_2$FeSi$_2$O$_7$, cf. $\alpha_{\text{GLSWT}} = 0.152$, and $\alpha_{\text{GLSWT+one-loop}} = 0.187$. This change is a direct consequence of the substantial renormalization of the energy $\omega_L(\mathbf{Q_m})$ of the longitudinal mode at the ZC. In fact, an accurate calculation that goes beyond the one-loop approximation estimates that the critical $\alpha_c$ required to close the gap $\omega_L(\mathbf{Q_m})$ for $\tilde{J}' = 0.1\tilde{J}$, and $\widetilde{\Delta} = \widetilde{\Delta}' = 1/3$ is around 0.158. In other words, the Hamiltonian parameters extracted from fitting the experiment with the GLSWT place Ba$_2$FeSi$_2$O$_7$ on the quantum paramagnetic side of the phase diagram shown in Fig. 1, which obviously contradicts the experimental evidence. In contrast, the set of parameters obtained from the GLSWT+one-loop correction ($\alpha_{\text{GLSWT+one-loop}}$) place the material at the magnetically ordered phase of exact phase diagram. Furthermore, the calculated ordered moment is very close to the measured value 2.95 $\mu_B$ (see Note 12 of the Supplementary Information for discussion of the reduction of the ordered moment). In general, non-linear corrections become increasingly important upon approaching the QCP and logarithmic corrections due to multi-loop vertex renormalizations become relevant very close to this point[28,31,53,54]. The fact that one-loop correction is enough to reproduce the spectrum of Ba$_2$FeSi$_2$O$_7$ indicates that this material is still far enough from that critical regime.

In summary, $Ba_2FeSi_2O_7$ provides a natural realization of a quasi-2D easy-plane antiferromagnet in the proximity of the QCP that signals the transition into the QPM phase. Previous examples of low-dimensional easy-plane quantum magnets in the proximity of this QCP were typically located on the quantum paramagnetic side of the quantum phase transition[17,20,21,23]. $Ba_2FeSi_2O_7$ then allows us to explain the strong decay and renormalization effects of the low-energy transverse and longitudinal modes of the AFM state. Furthermore, the distance to the O(2) QCP could be in principle controlled by chemical substitution, while the application of an in-plane magnetic field, that gaps out the transverse modes, can be used to control the decay rate of the longitudinal mode.

Here we have used the INS data of $Ba_2FeSi_2O_7$ as a platform to test a loop expansion based on an SU(3) spin wave theory[17,18,20,55], that captures the longitudinal and the transverse modes at the linear level. This loop expansion, that generalizes the well-known $1/S$-expansion of the SU(2) spin wave theory, allows us to reproduce the measured width and renormalization of the longitudinal and transverse modes near the zone center by just including a one-loop correction. Small discrepancies near the zone boundary are attributed to limitations of the effective low-energy $S = 1$ model that we adopted for this work.

The loop expansion that we have described in this manuscript provides a general scheme for treating quantum magnets with more than one type of low-energy mode. In general, quantum magnets that exhibit low-energy modes with $N - 1$ different "flavors" can be treated semi-classically using an SU($N$) spin wave theory. The parameter of the semi-classical expansion is the number of loops in the Feynman diagrams that contribute to the single-particle propagator.

**Methods**

**Sample preparation**

A single crystal of $Ba_2FeSi_2O_7$ was grown using an optical floating zone melting method[44]. Polycrystalline $Ba_2FeSi_2O_7$ feed-rods were prepared using the solid-state reaction method. The stoichiometric powders of $BaCO_3$ and $Fe_2O_3$, and $SiO_2$ were mixed, ground, pelletized and sintered with intermediate heating in a gas atmosphere. $Ba_2FeSi_2O_7$ single crystal was grown using a floating zone furnace in the same gas environment.

**Inelastic neutron scattering measurement**

Inelastic neutron scattering measurements were performed using the cold neutron triple-axis spectrometer (CTAX) at the High Flux Isotope Reactor (HFIR) and the hybrid spectrometer (HYSPEC) at the Spallation Neutron Source (SNS) at Oak Ridge National Laboratory[45]. A 2.15 g single crystal was aligned with the $(H, H, L)$ and $(H, 0, L)$ in the horizontal scattering plane for CTAX and HYSPEC experiments. A liquid helium cryostat was used to control temperature. At CTAX, the initial neutron energy was selected using a PG (002) monochromator, and the final neutron energy was fixed to $E_f$= 3.0 meV by a PG (002) analyzer. The horizontal collimation was

guide-open-40'-120', which provides an energy resolution with full width half maximum (FWHM)=0.1 and 0.18 meV for $\Delta E$=0 and 2.5 meV, respectively. For the HYSPEC experiment, $E_i$=9 meV and a Fermi chopper frequency of 300 Hz were used, which provides an energy resolution of FWHM=0.28 meV and 0.19 meV at $\Delta E$=0 and 2.5 meV, respectively. Measurements were performed at $T$=1.6 K and 90 K by rotating the sample from -50° to 170° with 1° steps. Data was symmetrized over positive and negative $H$ and integrated over $K$=[-0.1, 0.1] and $L$=[0.4, 0.6]. In Fig. 3a, there appears to be quasi-elastic scattering below 0.5 meV in low **Q**-region. This scattering arises from the incompletely blocked direct beam due to the oscillating collimator. All of data sets were reduced and analyzed using MANTID[56] and DAVE[57].

**Data availability**

The data sets generated during and/or analyzed during the current study are available from the corresponding authors on reasonable request.

**Code availability**

The codes used to generate the results in this work are available from the corresponding authors on reasonable request.

**References**

1. Lifshitz EM, Pitaevskii LP. Statistical Physics Part 2. Pergamon Press, Oxford (1980).
2. Pitaevskii LP. Properties of the spectrum of elementary excitations near the disintegration threshold of the excitations. *Sov Phys JETP* **9**, 830–837 (1959).
3. Woods ADB, Cowley RA. Structure and excitations of liquid helium. *Reports on Progress in Physics* **36**, 1135–1231 (1973).
4. Stone MB, Zaliznyak IA, Hong T, Broholm CL, Reich DH. Quasiparticle breakdown in a quantum spin liquid. *Nature* **440**, 187–190 (2006).
5. Hong T, *et al.* Field induced spontaneous quasiparticle decay and renormalization of quasiparticle dispersion in a quantum antiferromagnet. *Nat Commun* **8**, 15148 (2017).
6. Ma J, *et al.* Static and Dynamical Properties of the Spin-1/2 Equilateral Triangular-Lattice Antiferromagnet $Ba_3CoSb_2O_9$ *Phys Rev Lett* **116**, 087201 (2016).
7. Masuda T, Zheludev A, Manaka H, Regnault LP, Chung JH, Qiu Y. Dynamics of Composite Haldane Spin Chains in IPA-$CuCl_3$. *Phys Rev Lett* **96**, 047210 (2006).
8. Plumb KW, *et al.* Quasiparticle-continuum level repulsion in a quantum magnet. *Nat Phys* **12**, 224-229 (2016).
9. Zhitomirsky ME, Chernyshev AL. Colloquium: Spontaneous magnon decays. *Rev Mod Phys* **85**, 219–242 (2013).
10. Kim T, Park K, Leiner JC, Park J-G. Hybridization and Decay of Magnetic Excitations in Two-Dimensional Triangular Lattice Antiferromagnets. *J Phys Soc Jpn* **88**, 081003 (2019).
11. Chernyshev AL, Zhitomirsky ME. Spin waves in a triangular lattice antiferromagnet: Decays, spectrum renormalization, and singularities. *Phys Rev B* **79**, 144416 (2009).
12. Kamiya Y, *et al.* The nature of spin excitations in the one-third magnetization plateau phase of $Ba_3CoSb_2O_9$. *Nat Commun* **9**, 2666 (2018).
13. Leiner JC, *et al.* Magnetic excitations in the bulk multiferroic two-dimensional triangular lattice antiferromagnet (Lu,Sc)$FeO_3$. *Phys Rev B* **98**, 134412 (2018).

14. Park P, *et al.* Momentum-Dependent Magnon Lifetime in the Metallic Noncollinear Triangular Antiferromagnet $CrB_2$. *Phys Rev Lett* **125**, 027202 (2020).
15. Thompson JD, McClarty PA, Prabhakaran D, Cabrera I, Guidi T, Coldea R. Quasiparticle Breakdown and Spin Hamiltonian of the Frustrated Quantum Pyrochlore $Yb_2Ti_2O_7$ in a Magnetic Field. *Phys Rev Lett* **119**, 057203 (2017).
16. Zheng W, Fjærestad JO, Singh RRP, McKenzie RH, Coldea R. Anomalous Excitation Spectra of Frustrated Quantum Antiferromagnets. *Phys Rev Lett* **96**, 057201 (2006).
17. Kohama Y, *et al.* Thermal Transport and Strong Mass Renormalization in $NiCl_2$-$4SC(NH_2)_2$. *Phys Rev Lett* **106**, 037203 (2011).
18. Muniz RA, Kato Y, Batista CD. Generalized spin-wave theory: Application to the bilinear–biquadratic model. *Prog Theor Exp Phys* **2014**, 083I001 (2014).
19. Bai X, *et al.* Hybridized quadrupolar excitations in the spin-anisotropic frustrated magnet $FeI_2$. *Nat Phys*, (2021).
20. Zapf V, Jaime M, Batista CD. Bose-Einstein condensation in quantum magnets. *Rev Mod Phys* **86**, 563–614 (2014).
21. Zapf VS, *et al.* Bose-Einstein Condensation of *S*=1 Nickel Spin Degrees of Freedom in $NiCl_2$-$4SC(NH_2)_2$. *Phys Rev Lett* **96**, 077204 (2006).
22. Zhang Z, Wierschem K, Yap I, Kato Y, Batista CD, Sengupta P. Phase diagram and magnetic excitations of anisotropic spin-one magnets. *Phys Rev B* **87**, 174405 (2013).
23. Zvyagin SA, *et al.* Magnetic Excitations in the Spin-1 Anisotropic Heisenberg Antiferromagnetic Chain System $NiCl_2$-$4SC(NH_2)_2$. *Phys Rev Lett* **98**, 047205 (2007).
24. Pekker D, Varma CM. Amplitude/Higgs Modes in Condensed Matter Physics. *Annu Rev Condens Matter Phys* **6**, 269-297 (2015).
25. Chubukov AV, Sachdev S, Ye J. Theory of two-dimensional quantum Heisenberg antiferromagnets with a nearly critical ground state. *Phys Rev B* **49**, 11919–11961 (1994).
26. Sachdev S. Universal relaxational dynamics near two-dimensional quantum critical points. *Phys Rev B* **59**, 14054–14073 (1999).
27. Gazit S, Podolsky D, Auerbach A. Fate of the Higgs Mode Near Quantum Criticality. *Phys Rev Lett* **110**, 140401 (2013).
28. Podolsky D, Auerbach A, Arovas DP. Visibility of the amplitude (Higgs) mode in condensed matter. *Phys Rev B* **84**, 174522 (2011).
29. Podolsky D, Sachdev S. Spectral functions of the Higgs mode near two-dimensional quantum critical points. *Phys Rev B* **86**, 054508 (2012).
30. Rose F, Léonard F, Dupuis N. Higgs amplitude mode in the vicinity of a (2+1)-dimensional quantum critical point: A nonperturbative renormalization-group approach. *Phys Rev B* **91**, 224501 (2015).
31. Kulik Y, Sushkov OP. Width of the longitudinal magnon in the vicinity of the O(3) quantum critical point. *Phys Rev B* **84**, 134418 (2011).
32. Scammell HD, Sushkov OP. Asymptotic freedom in quantum magnets. *Phys Rev B* **92**, 220401 (2015).
33. Sinner A, Hasselmann N, Kopietz P. Spectral function and quasiparticle damping of interacting bosons in two dimensions. *Phys Rev lett* **102**, 120601 (2009).
34. Jain A, *et al.* Higgs mode and its decay in a two dimensional antiferromagnet. *Nat Phys* **13**, 633-637 (2017).
35. Anderson PW. An approximate quantum theory of the antiferromagnetic ground state. *Physical Review* **86**, 694 (1952).

36. Auerbach A. *Interacting electrons and quantum magnetism*. Springer-Verlag (1994).
37. Holstein T, Primakoff H. Field dependence of the intrinsic domain magnetization of a ferromagnet. *Physical Review* **58**, 1098 (1940).
38. Kubo R. The spin-wave theory of antiferromagnetics. *Physical Review* **87**, 568 (1952).
39. Manousakis E. The spin-½ Heisenberg antiferromagnet on a square lattice and its application to the cuprous oxides. *Rev Mod Phys* **63**, 1 (1991).
40. Mattis DC. *The theory of magnetism I: Statics and Dynamics*. Springer Science & Business Media (2012).
41. Chubukov AV, Sachdev S, Senthil T. Large-$S$ expansion for quantum antiferromagnets on a triangular lattice. *Journal of Physics: Condensed Matter* **6**, 8891–8902 (1994).
42. Sidney C. *Aspects of Symmetry*. Cambridge (1985).
43. Mai TT, *et al.* Terahertz spin-orbital excitations in the paramagnetic state of multiferroic $Sr_2FeSi_2O_7$ *Phys Rev B* **94**, (2016).
44. Jang T-H, *et al.* Physical properties of a quasi-two-dimensional square lattice antiferromagnet $Ba_2FeSi_2O_7$. Preprint at https://arxiv.org/abs/2108.00999 (2021).
45. Winn B, *et al.* Recent progress on HYSPEC, and its polarization analysis capabilities. *EPJ Web of Conferences* **83**, 03017 (2015).
46. Romhányi J, Penc K. Multiboson spin-wave theory for $Ba_2CoGe_2O_7$: A spin-3/2 easy-plane Néel antiferromagnet with strong single-ion anisotropy. *Phys Rev B* **86**, 174428 (2012).
47. Papanicolaou N. Unusual phases in quantum spin-1 systems. *Nuclear Physics B* **305**, 367-395 (1988).
48. Arovas DP, Auerbach A. Functional integral theories of low-dimensional quantum Heisenberg models. *Phys Rev B* **38**, 316–332 (1988).
49. Ghioldi EA, *et al.* Dynamical structure factor of the triangular antiferromagnet: Schwinger boson theory beyond mean field. *Phys Rev B* **98**, 184403 (2018).
50. Zhang S-S, Ghioldi EA, Kamiya Y, Manuel LO, Trumper AE, Batista CD. Large-$S$ limit of the large-$N$ theory for the triangular antiferromagnet. *Phys Rev B* **100**, 104431 (2019).
51. Gnutzmann S, Kus M. Coherent states and the classical limit on irreducible representations. *Journal of Physics A: Mathematical and General* **31**, 9871 (1998).
52. Mourigal M, Fuhrman WT, Chernyshev AL, Zhitomirsky ME. Dynamical structure factor of the triangular-lattice antiferromagnet. *Phys Rev B* **88**, 094407 (2013).
53. Affleck I, Wellman GF. Longitudinal modes in quasi-one-dimensional antiferromagnets. *Phys Rev B* **46**, 8934 (1992).
54. Nohadani O, Wessel S, Haas S. Quantum phase transitions in coupled dimer compounds. *Phys Rev B* **72**, 024440 (2005).
55. Matsumoto M. Electromagnon as a Probe of Higgs (Longitudinal) Mode in Collinear and Noncollinear Magnetically Ordered States. *J Phys Soc Jpn* **83**, 084704 (2014).
56. Taylor, *et al.* Mantid, A high performance framework for reduction and analysis of neutron scattering data. *Bulletin of the American Physical Society* **57**, (2012).
57. Azuah RT, *et al.* DAVE: A comprehensive software suite for the reduction, visualization, and analysis of low energy neutron spectroscopic data. *J Res Natl Inst Stan Technol* **114**, 341 (2009).

**Acknowledgments**

We thank Shang-Shun Zhang, Jie Xing, and Andrew F. May for useful discussions and Choongjae Won for helping with sample growth. This work was supported by the U.S. Department of Energy, Office of Science, Basic Energy Sciences, Materials Science and Engineering Division. This research used resources at the High Flux Isotope Reactor and Spallation Neutron Source, DOE Office of Science User Facilities operated by the Oak Ridge National Laboratory (ORNL). Access to MACS was provided by the Center for High Resolution Neutron Scattering, a partnership between the National Institute of Standards and Technology and the National Science Foundation under Agreement No. DMR-1508249. The work at Max Planck POSTECH/Korea Research Initiative was supported by Nano Scale Optomaterials and Complex Phase Materials (2016K1A4A4A01922028), through the National Research Foundation (NRF) funded by MSIP of Korea. The work at Rutgers University was supported by the DOE under Grant No. DOE: DE-FG02-07ER46382.

**Author contributions**

S.H.D. and A.D.C. conceived the project, which was supervised by C.D.B. and A.D.C.. T.H.J., S.W.C., and J.-H.P. provided single crystals. T.H.J. measured physical properties. S.H.D., T.J.W., T.H., V.O.G., J.A.R. and A.D.C. performed INS experiments. S.H.D., T.J.W., and A.D.C. analyzed the neutron data. H.Z. and C.D.B constructed theoretical model and calculations. S.-H.D., H.Z., C.D.B., and A.D.C. wrote the manuscript with input from all authors.

**Additional information**

Supplementary Information are available.

**Competing interests**

The authors declare that they have no competing interests.

**Figures and Tables**

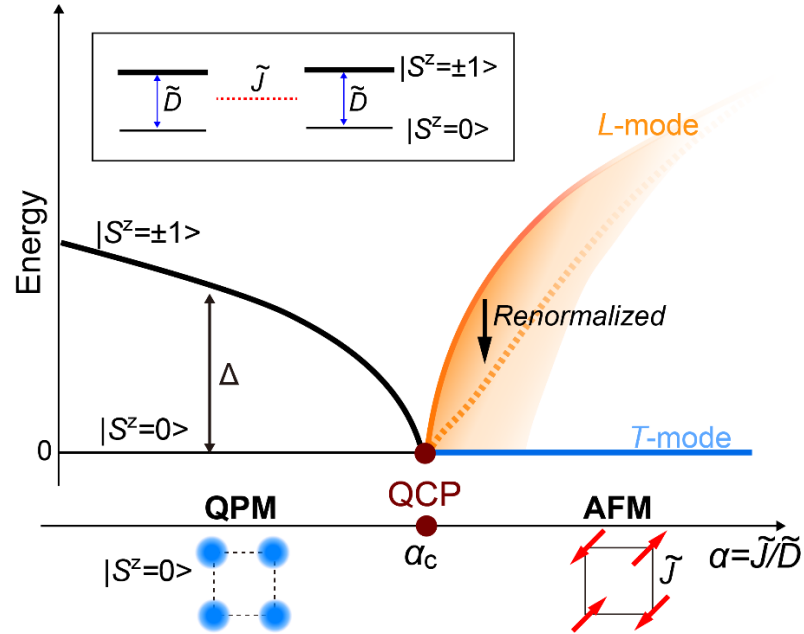

**Fig. 1 Schematic diagrams near the quantum critical point.**
Schematic phase diagram illustrates the O(2) quantum critical point (QCP) between the antiferromagnetic (AFM) state and the quantum paramagnet (QPM) as a function of $\alpha = \tilde{J}/\tilde{D}$ ($\tilde{J}$ is a Heisenberg exchange and $\tilde{D}$ is a easy-plane single-ion anisotropy of effective $S$=1). The low-energy excitations of the QPM are two degenerate $S^z=\pm 1$ modes (black line) with a gap, Δ, which closes at the QCP. The spontaneous U(1) symmetry breaking leads to a gapless magnon or transverse mode ($T$-mode), indicated with a blue line, which is accompanied by a gapped longitudinal mode ($L$-mode) indicated with the orange line. Near the QCP, the energy and the lifetime of the $L$-mode are strongly renormalized (dashed orange line) due to the decay into the continuum of two transverse modes (shaded orange region).

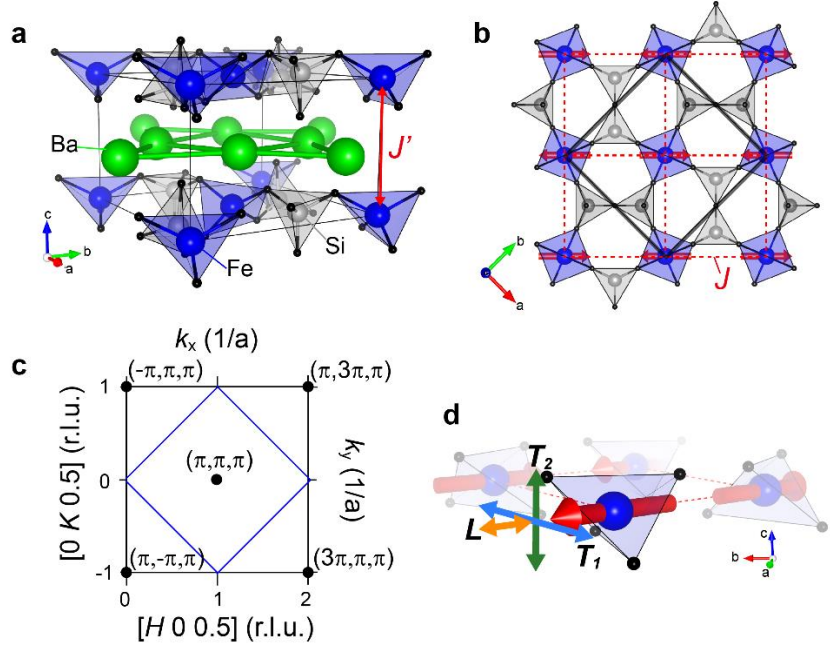

**Fig. 2 Crystal and magnetic structure of Ba$_2$FeSi$_2$O$_7$.**
**a** Crystal structure of Ba$_2$FeSi$_2$O$_7$. Ba atoms separate layers composed of FeSi$_2$O$_7$, rendering a quasi-two-dimensional structure. **b** In the FeSi$_2$O$_7$ layer, FeO$_4$ tetrahedra are connected via SiO$_4$ polyhedra, and the adjacent two Fe$^{2+}$ atoms are exchange coupled by two oxygen ligands. The red dashed line indicates the exchange pathway $J$ within two-dimensional square spin network. The interlayer coupling $J'$ is found here to be much weaker than $J$. Red arrows indicate the moment direction in the collinear AFM phase as determined in Ref.[44]. The black solid line indicates the chemical unit cell. **c** $HK$-reciprocal space with $L=0.5$ in the tetragonal structure ($P\bar{4}2_1m$). The blue solid line and the black circle indicate the Brillouin zone and zone center, respectively. The coordinates ($H, K, L$) of the reciprocal lattice of the origin lattice are related to ($k_x, k_y, k_z$) of the magnetic lattice formed by the Fe$^{2+}$ atoms through $k_x = \pi(H - K)$, $k_y = \pi(H + K)$, and $k_z = 2\pi L$. **d** Illustration of the spin fluctuation modes. $T_1$ and $T_2$ indicate transverse fluctuation in the $ab$-plane and out-of the plane, respectively. $L$ indicates longitudinal fluctuation of spin.

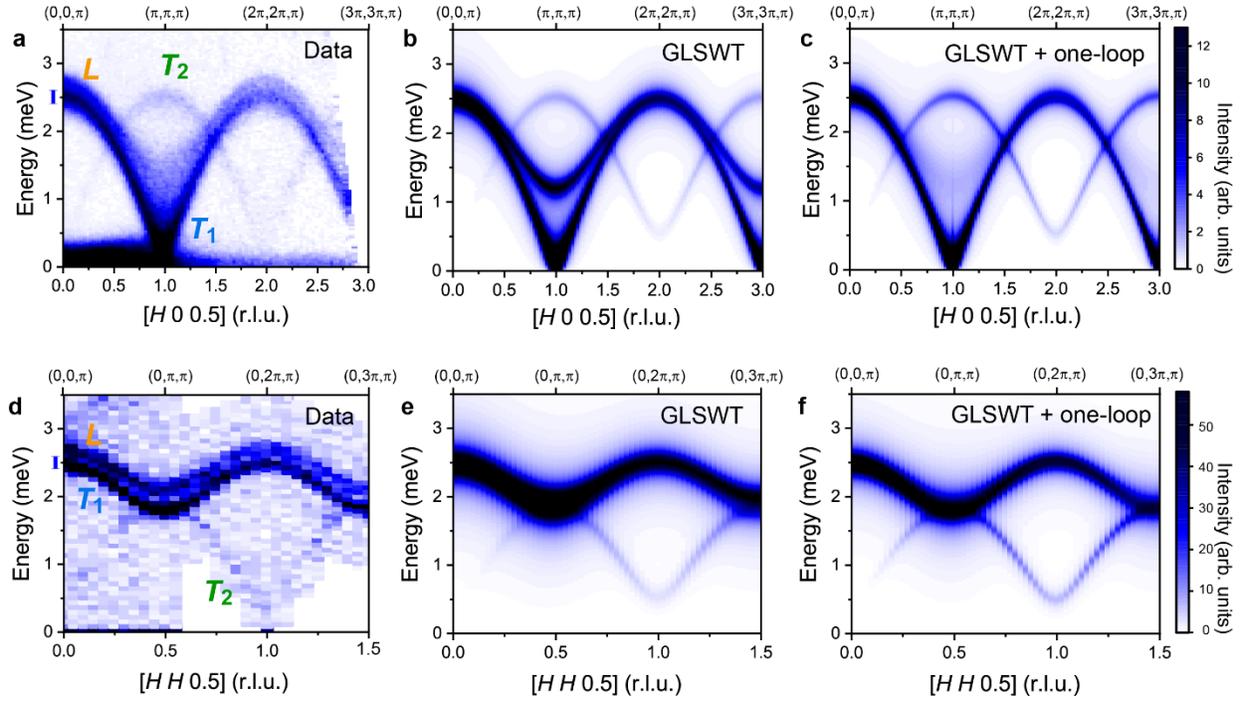

**Fig. 3 Inelastic neutron scattering of Ba$_2$FeSi$_2$O$_7$**

**a** Contour map of the inelastic neutron scattering (INS) data as a function of energy and momentum transfer along the [$H$, 0, 0.5] direction measured at $T$=1.6 K ($< T_N$) using the HYSPEC time-of-flight spectrometer at SNS. **d** Contour map of the INS data as a function of energy and momentum transfer along [$H,H,0.5$] direction measured at $T$=1.4 K ($< T_N$) using the cold Neutron Triple-Axis spectrometer (CTAX) at HFIR. The instrumental resolutions at energy=2.5 meV for each instrument are indicated with blue bars along the $y$-axis in **a** and **d**. The two transverse modes and the longitudinal mode are labeled with $T_1$, $T_2$, and $L$, respectively. **b**, **c**, **e**, and **f** INS intensities calculated by the generalized linear spin wave theory (GLSWT) and GLSWT plus one-loop corrections (GLSWT+one-loop) with the parameter sets $\mathcal{A}$ and $\mathcal{B}$ given in Table. 1, respectively. The instrumental resolution of HYSPEC and CTAX was modeled in the calculated spectra using a Lorentzian function.

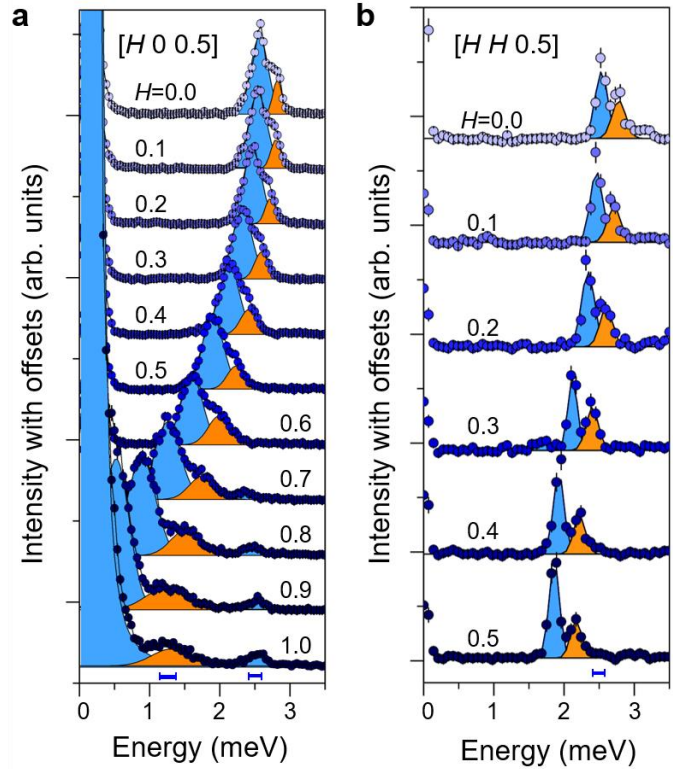

20
**Fig. 4 Detailed line-cuts of INS spectra.**
**a** Constant momentum cuts at points along the [$H$, 0, 0.5]-direction measured using HYSPEC at SNS, integrated over $H=[H-0.05, H+0.05]$ at selected $H$, $K=[-0.1, 0.1]$, and $L=[0.4, 0.6]$. **b** Constant momentum cuts at points along the [$H$, $H$, 0.5]-direction measured using CTAX at HFIR. Blue bars at the bottom of the panels indicate the instrumental resolutions for HYSPEC and CTAX at the proximate energy transfers. The blue and orange shaded regions are the results of fitting Gaussian line shapes to transverse ($T_1$, $T_2$) and longitudinal ($L$) modes, respectively.

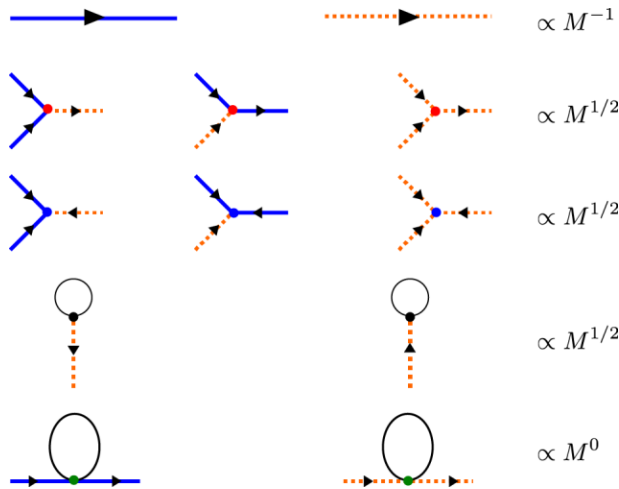

**Fig. 5 Basic ingredients of the perturbative field theory in $1/M$ for Ba$_2$FeSi$_2$O$_7$.**
Solid (dash) lines represent the bare propagator of the transverse (longitudinal) boson. The symmetry-allowed cubic vertices are shown on the second and third lines. The red (blue) dot represents a decay (sink) vertex. The cubic-linear vertices are listed on the fourth line. The last line represents the normal vertex $V_{\alpha\alpha}^{(4,N)}$ from $\mathcal{H}^{(4)}$.

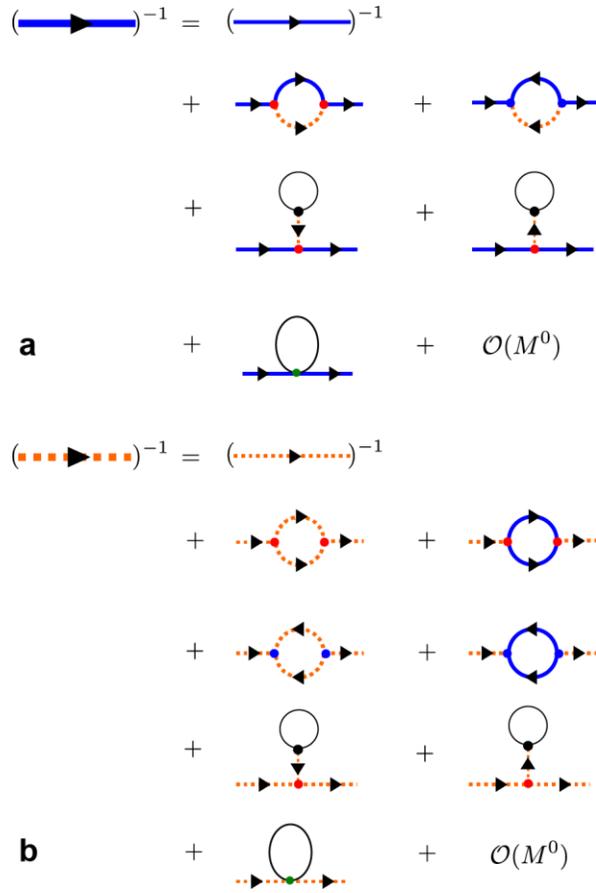

**Fig. 6 Diagrammatic representation of the Dyson equation.**
**a** One-loop diagrams that contribute up to the order $M^0$ for the transverse boson. **b** One-loop diagrams that contribute up to the order $M^0$ for the longitudinal boson. The dressed propagator is denoted by a thick line, whereas the bare propagator is denoted by a thin line.

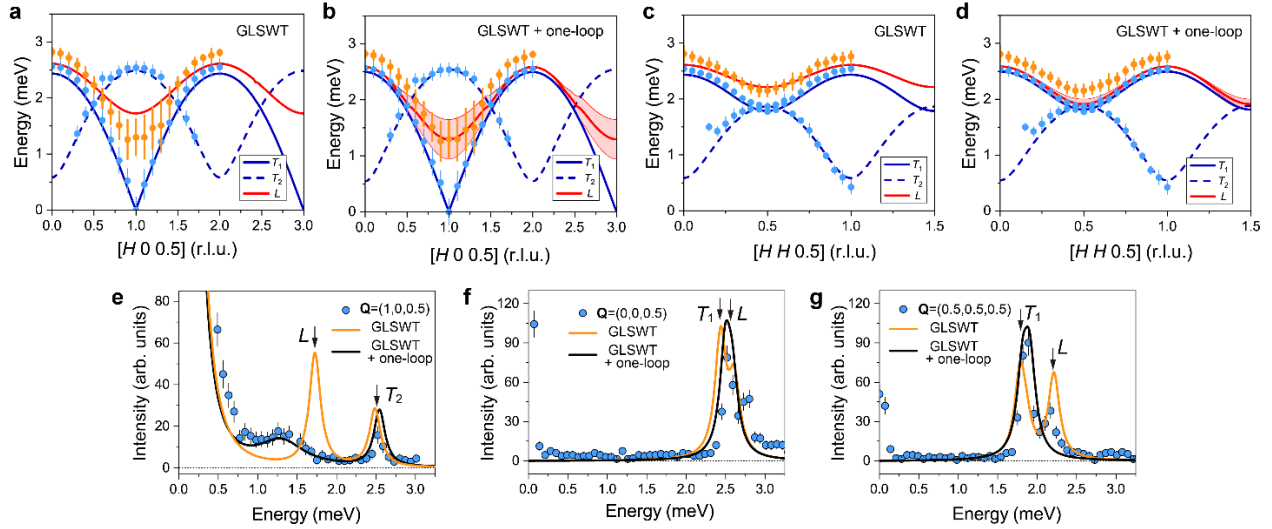

Fig. 7 **Comparison between measured and calculated spectrum.**
Comparison of the measured and calculated dispersion along the $[H, 0, 0.5]$ (**a**, **b**) and $[H, H, 0.5]$ (**c**, **d**) directions. In all panels of this figure, the theoretical results are obtained for the parameter set $\mathcal{B}$ in Table 1. **a-d** Blue and orange filled circles indicate the measured transverse and longitudinal modes, obtained from the Gaussian fitting to the data shown in Fig. 4a. Dots and error bars indicate peak centers and full width at half maxima (FWHM) of the observed modes, respectively. Lines indicate the calculated dispersions obtained from the GLSWT and GLSWT+one-loop corrections. The red shaded region in **b** and **d** depict the decay (line-broadening) of the longitudinal mode given by the one-loop corrections. **e-g** Comparison between the measured (blue dots) and calculated (orange and black lines) INS intensities at three high-symmetric **Q**-points at (1, 0, 0.5), (0, 0, 0.5), and (0.5, 0.5, 0.5). All the experimental data were measured using CTAX with fixed $E_f$=3 meV. For GLSWT, two transverse and longitudinal modes are denoted with $T_1$, $T_2$, and $L$.

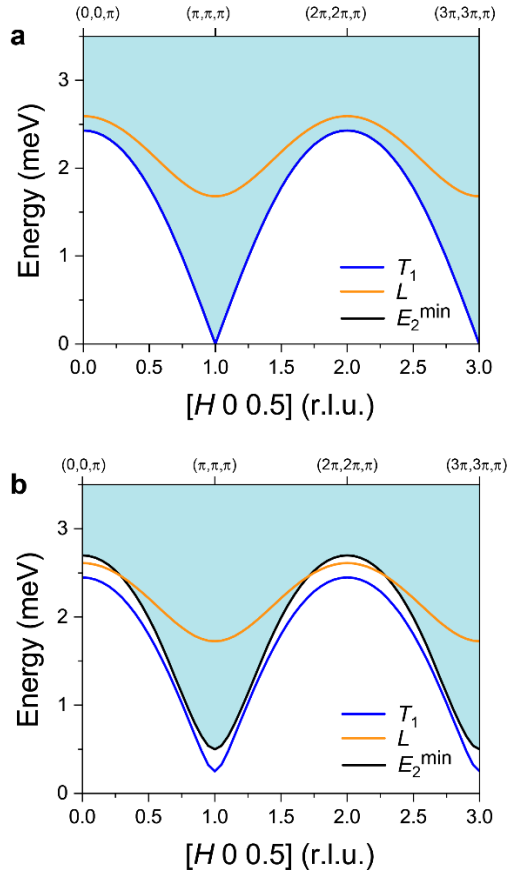

**Fig. 8 Kinematic constraints for the decay of the longitudinal mode.**
The blue (orange) curve shows the calculated transverse (longitudinal) band dispersions along [$H$, 0, 0.5] with the GLSWT (using parameters set $\mathcal{B}$ in Table 1). The light blue-shaded areas indicate the two-transverse mode continuum, whose lower edge is indicated with a black solid line ($E_2^{\min}$).
**a** Results of the effective $S = 1$ model. **b** Same as **a** but for a gapped branch of transverse modes (an ad hoc gap has been added to Eq. (21)).

**Table 1 | Parameter sets of GLSWT and GLSWT+one-loop models.**
The parameters of the effective $S = 1$ model extracted by fitting the Gaussian-peak centers of the experimental dispersion with the GLSWT and GLSWT + one-loop calculated energies at the zone center $\mathbf{Q_m}$=(1, 0, 0.5) In both cases, we assume $\tilde{J}' = 0.1\tilde{J}$, and $\tilde{\Delta} = \tilde{\Delta}' = 1/3$, i.e. $\Delta = \Delta' = 1$ for the $S = 2$ model (Heisenberg model without exchange anisotropy). The parameter set is referred to by its label ($\mathcal{A}$ or $\mathcal{B}$) in the text.

| Theory | Label | $\tilde{J}$ (meV) | $\tilde{D}$ (meV) |
| --- | --- | --- | --- |
| GLSWT | $\mathcal{A}$ | 0.245 (7) | 1.61 (6) |
| GLSWT + one-loop | $\mathcal{B}$ | 0.266 (6) | 1.42 (4) |

# Supplementary Information

## Supplementary Note 1. Single-ion state of $Fe^{2+}$ in $Ba_2FeSi_2O_7$

Recent terahertz spectroscopy and X-ray absorption spectroscopy (XAS) on $A_2FeSi_2O_7$ (A=Sr and Ba) revealed that the considerable tetragonal distortion of FeO$_4$-tetrahedra (Sr: 17% and Ba: 26% z-compression from cubic) with large spin-orbit coupling of $Fe^{2+}$ ($\lambda \sim 20$ meV) induces significant easy-plane single-ion anisotropy in the system[1,2]. Following the description for the spin-orbital states of $Fe^{2+}$ in a tetrahedral environment, in $Ba_2FeSi_2O_7$ the large single-ion anisotropy determines the ground state from the multiplet of $Fe^{2+}$ ion. Starting with the free ion, $^5D$ ($L = 2, S = 2$), the tetrahedral crystal field ($\Delta_{Td}$) and tetragonal distortion ($\delta_{Tetra}$) leave an A-manifold with 5 levels (see Supplementary Figure 1) with hybridized $L^z$ and $S^z$ states. When $\Delta_{Td}, \delta_{Tetra} \gg \lambda$, it leads to a pure spin $S = 2$ quintet (see Supplementary Figure 1a). Introducing a single-ion term $D(S^z)^2$ with easy-plane anisotropy, lifts the degeneracy of the quintet into levels with $S^z = 0$ (singlet), $S^z = \pm 1$ (doublet), and $S^z = \pm 2$ (doublet) where the energy-splitting between the states is given $D$ and $3D$. Supplementary Figure 1b shows the inelastic neutron scattering measured at $T$=90 K, $\sim 7 \ast \Theta_{CW}$ (powder averaged Curie-Weiss temperature $\Theta_{CW} \sim$ -12.8 K) where the single-ion physics dominates[1]. In the spectra, two flat excitations are visible at $dE \sim 1.32$ meV and 3.9 meV which indicate transitions between levels as follows $|S^z = 0\rangle \rightarrow |S^z = \pm 1\rangle$ and $|S^z = \pm 1\rangle \rightarrow |S^z = \pm 2\rangle$. Note that the transitions between $|S^z = 0\rangle$ and $|S^z = \pm 2\rangle$ are forbidden by dipole selection rules. At finite temperature, the thermal population of the spin states determines the effective spin of the system. In the low temperature region where $T \ll 3 \ast D$ ($\sim$45 K), the $|S^z = \pm 2\rangle$ states are depopulated and the system can then be treated as an effective $S = 1$.

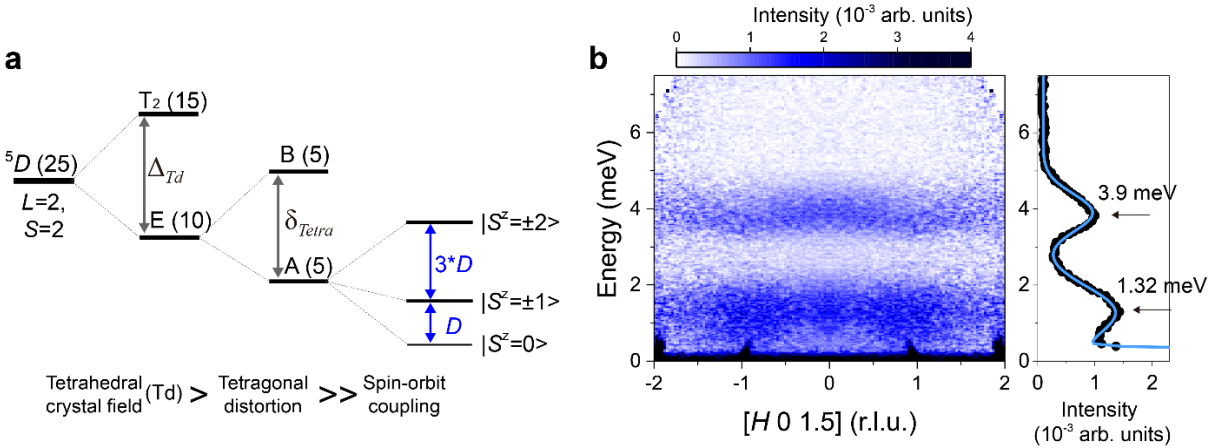

**Supplementary Figure 1. Orbital configuration and single-ion excitation of $Ba_2FeSi_2O_7$**
**a** Orbital energy states of $Fe^{2+}$ with a tetrahedral crystal field ($\Delta_{Td}$), tetragonal distortion ($\delta_{Tetra}$), and spin orbit coupling ($\lambda$). **b** The left panel shows inelastic neutron scattering data measured at $T$=90 K, symmetrized over negative and positive $H$ and integrated over $L$=[0.9, 2.1] and $K$=[-0.1, 0.1]. The integrated scattering intensity over $H$=[-2, 2] is shown in the right panel. The two peaks were fitted with Gaussian functions (solid blue line). Arrows indicate the peak centers at 1.32 meV and 3.9 meV.

**Supplementary Note 2. Analysis of a possible spin gap**

To check for evidence of a gap in the spin wave dispersion at the ZC, the low energy inelastic neutron scattering was investigated using CTAX at HFIR with $E_f$=3 meV. The energy resolution of the instrument (FWHM~0.101 meV for elastic scattering) poses a challenge to directly extract a small gap due to the large scattering near the magnetic Bragg reflection at $\mathbf{Q_m}$. As an alternative approach, we examine the spin wave dispersions near the ZC and compare the calculated values with and without a small gap. The measured dispersion was obtained by fitting with a resolution convoluted Gaussian function to constant momentum transfer scans for $0.85 \leq H \leq 1.15$. The extracted magnon dispersion is displayed as the contour plot along with the calculated spin waves with $\Delta_{\text{gap}} = 0$ and 0.25 meV in the Supplementary Figure 2a. The gapless Goldstone mode ($\Delta_{\text{gap}}$=0) has linear dispersion emanating from the ZC, whereas the gaped tranverse mode has a quadratic dispersion near the ZC providing a possible means of distinguising a gaped spectrum from an gapless one. To find the best description of the dispersion near the ZC, the deviation between the data and calculated dispersion is defined as $(E_{exp.} - E_{calc.})^2$. The sum of this deviation is presented in Supplementary Figure 2b as a function of the gap energy. The deviation has a minimum at $\Delta_{\text{gap}} \sim 0.25$ meV, suggesting a gap in spin wave spectrum of $Ba_2FeSi_2O_7$.

The spin gap can also be extracted by extrapolating the magnetic field dependence of the $T_1$ transverse mode. The field-dependent low energy inelastic neutron scattering was measured using the Multi-Axis Crystal Spectrometer (MACS) at NCNR with $E_f$=3 meV. The constant $\mathbf{Q}$-linecuts were obtained at ZC with applied fields, $H$//[1 0 0], at 0, 1, 2, 3, 4, and 5 T. Supplementary Figure 2c shows the field-evolution of the spin excitation spectrum at the ZC. Each constant $\mathbf{Q}$-linecuts were fitted to a Gaussian function to parameterize the $T_1$ transverse modes. The $T_1$ transverse mode becomes gapped by the Zeeman energy in a magnetic field, and the gap increases with the magnetic field. For a gapless system the linear extrapolation of the $T_1$-modes approaches zero energy as the field goes to zero, whereas the dispersion relation of the gapped magnon in the transverse field is described by $\omega_{mag} \propto \sqrt{\Delta_{gap}^2 + (c)^2 H^2}$, where $\Delta_{gap}$ and $c$ can be fit to the experimental data[4]. The measured $T_1$-mode dispersions were fitted to the gapless linear and gaped parabola functions, which give $\chi^2$ values with 0.15 and 0.02, respectively. The lower $\chi^2$ result of the parabola function indicates the presence of a spin gap. The resulting value of the spin gap, $\Delta_{gap}$=0.18 (2) meV, is close to the value obtained from fitting to the low energy dispersion, implying that both analyses consistently indicate the presence of a small gap in the spin excitation spectrum of $Ba_2FeSi_2O_7$.

**Supplementary Note 3. Magnetic susceptibility with angular field-dependence**

The angular dependent magnetic susceptibility provides evidence for the single-ion anisotropy necessary to produce a gap in the spectrum. We measured this angular dependence on a $Ba_2FeSi_2O_7$ crystal that was aligned using X-ray Laue diffraction. Magnetization was measured using vibrating sample magnetometry (VSM) implemented in a Quantum Design physical properties measurement system (PPMS-Dynacool).

Supplementary Figure 3a shows the magnetic susceptibilities, χ(T), measured with three diffrent magnetic field directions in the *ab*-plane, parallel to [1, 0, 0], [1, 1, 0], and [0, 1, 0] of crystal axes. The three χ(T) show isotropic behavior above the Néel temperature ($T_N$=5.2 K), collapsing in a single line. Noticeably, those become anisotropic with magnetic order below the $T_N$, having a weak easy-axis anisotropy along [1, 1, 0]-direction. As described in the main text, the tetragonal symmetry with *S*=2 spin allows for a single-ion anisotropy term $\mathcal{H}_A = A \sum_i [(S_i^x)^4 + (S_i^y)^4]$ (*A*>0 for easy-axis along [1, 1, 0] and [1, -1, 0]), which can induce easy-axis anisotropy in the *ab*-plane. This anisotropy term can generate a gap in the spin wave spectrum as described in Note 2.

**Supplementary Note 4. Spin wave dispersion along *L*-direction**

Supplementary Figure 4 shows INS data along the *L*-direction for [*H*, 0, *L*] with *H*=0 and 1, measured using HYSPEC spectrometer. The spin excitations are weakly dispersing along *L* indicating quasi-two-dimensional behavior due to the relatively weak inter-layer coupling. As shown, the acoustic magnon ($T_1$) has a bandwidth of 0.5 meV, and the $T_2$ and *L* modes are almost completely flat along the *L* direction. The *L*-dependence of the spin excitations is reproduced by the GLSWT and GLSWT + one loop correction calculations with $\widetilde{J}'=\widetilde{J}/10$.

**Supplementary Note 5. Generalized spin wave approach**

In the local reference frame, the spin and quadrupolar operators can be expanded in $1/M$ as:

$$s_{\mathbf{r}}^{\mu} = M\tilde{S}_{00}^{\mu} + \sqrt{M} \sum_{\alpha=\pm 1} \left(\tilde{S}_{\alpha 0}^{\mu} b_{\mathbf{r},\alpha}^{\dagger} + h.c.\right)$$
$$+ \sum_{\alpha,\beta=\pm 1} \left(\tilde{S}_{\alpha\beta}^{\mu} - \tilde{S}_{00}^{\mu}\delta_{\alpha\beta}\right) \tilde{b}_{\mathbf{r},\alpha}^{\dagger}\tilde{b}_{\mathbf{r},\beta} \quad (S1)$$
$$- \frac{1}{2\sqrt{M}} \sum_{\alpha=\pm 1}\sum_{\beta=\pm 1} (\tilde{S}_{\alpha 0}^{\mu} \tilde{b}_{\mathbf{r},\alpha}^{\dagger}\tilde{b}_{\mathbf{r},\beta}^{\dagger}\tilde{b}_{\mathbf{r},\beta} + h.c.) + \mathcal{O}(\frac{1}{M^{3/2}}),$$

$$(s_{\mathbf{r}}^z)^2 = 1 - M\tilde{\mathcal{A}}_{00} - \sqrt{M} \sum_{\alpha=\pm 1} (\tilde{\mathcal{A}}_{\alpha 0}\tilde{b}_{\mathbf{r},\alpha}^{\dagger} + h.c.)$$
$$- \sum_{\alpha,\beta=\pm 1} (\tilde{\mathcal{A}}_{\alpha\beta} - \tilde{\mathcal{A}}_{00}\delta_{\alpha\beta})\tilde{b}_{\mathbf{r},\alpha}^{\dagger}\tilde{b}_{\mathbf{r},\beta} \quad (S2)$$
$$+ \frac{1}{2\sqrt{M}} \sum_{\alpha=\pm 1}\sum_{\beta=\pm 1} (\tilde{\mathcal{A}}_{\alpha 0}\tilde{b}_{\mathbf{r},\alpha}^{\dagger}\tilde{b}_{\mathbf{r},\beta}^{\dagger}\tilde{b}_{\mathbf{r},\beta} + h.c.) + \mathcal{O}(\frac{1}{M^{3/2}}).$$

The expressions for the coefficients $A_{\mathbf{k},\alpha}$ and $B_{\mathbf{k},\alpha}$ of the quadratic Hamiltonian $\mathcal{H}^{(2)}$ Eq. (17) are:

$$A_{\mathbf{k},+1} = -8(x-1)x(2\tilde{J}+\tilde{J}') - (x-1)\tilde{D} \quad (S3)$$

$$+2(x(1+\tilde{\Delta})-1)\tilde{J}\gamma_\mathbf{k}^{xy}$$
$$+2(x(1+\tilde{\Delta}')-1)\tilde{J}'\gamma_\mathbf{k}^z,$$
$$B_{\mathbf{k},+1} = -2(x(\tilde{\Delta}-1)+1)\tilde{J}\gamma_\mathbf{k}^{xy}$$
$$-2(x(\tilde{\Delta}'-1)+1)\tilde{J}'\gamma_\mathbf{k}^z,$$
$$A_{\mathbf{k},-1} = -16(x-1)x(2\tilde{J}+\tilde{J}') - (2x-1)D$$
$$-2(1-2x)^2(\tilde{J}\gamma_\mathbf{k}^{xy} + \tilde{J}'\gamma_\mathbf{k}^z),$$
$$B_{\mathbf{k},-1} = 2(1-2x)^2(\tilde{J}\gamma_\mathbf{k}^{xy} + \tilde{J}'\gamma_\mathbf{k}^z)$$

**Supplementary Note 6. Cubic and cubic-linear vertices**

In this section, we derive the cubic and cubic-linear vertices given in Eq. (24) and Eq. (25). The cubic Hamiltonian has three contributions

$$\mathcal{H}^{(3)} = \mathcal{H}_{intra}^{(3)} + \mathcal{H}_{inter}^{(3)} + \mathcal{H}_D^{(3)}, \tag{S4}$$

with

$$\mathcal{H}_{intra}^{(3)} = \tilde{J} \sum_{\langle \mathbf{r},\mathbf{r}'\rangle,\nu} \sum_{\alpha,\beta=\pm 1} \{\sum_{\alpha'=\pm 1} a_\nu [2\tilde{S}_{\alpha\beta}^\nu \tilde{S}_{0\alpha'}^\nu \tilde{b}_{\mathbf{r}\alpha}^\dagger \tilde{b}_{\mathbf{r}\beta} \tilde{b}_{\mathbf{r}'\alpha'}]$$
$$-a_\nu [\tilde{S}_{0\alpha}^\nu \tilde{S}_{00}^\nu (\tilde{b}_{\mathbf{r}\beta}^\dagger \tilde{b}_{\mathbf{r}\beta} \tilde{b}_{\mathbf{r}\alpha} + 2\tilde{b}_{\mathbf{r}'\beta}^\dagger \tilde{b}_{\mathbf{r}'\beta} \tilde{b}_{\mathbf{r}\alpha})] + h.c.\}, \tag{S5}$$

$$\mathcal{H}_{inter}^{(3)} = \tilde{J}' \sum_{\langle \mathbf{r},\mathbf{r}'\rangle,\nu} \sum_{\alpha,\beta=\pm 1} \{\sum_{\alpha'=\pm 1} b_\nu [2\tilde{S}_{\alpha\beta}^\nu \tilde{S}_{0\alpha'}^\nu \tilde{b}_{\mathbf{r}\alpha}^\dagger \tilde{b}_{\mathbf{r}\beta} \tilde{b}_{\mathbf{r}'\alpha'}]$$
$$-b_\nu [\tilde{S}_{0\alpha}^\nu \tilde{S}_{00}^\nu (\tilde{b}_{\mathbf{r}\beta}^\dagger \tilde{b}_{\mathbf{r}\beta} \tilde{b}_{\mathbf{r}\alpha} + 2\tilde{b}_{\mathbf{r}'\beta}^\dagger \tilde{b}_{\mathbf{r}'\beta} \tilde{b}_{\mathbf{r}\alpha})] + h.c.\}, \tag{S6}$$

$$\mathcal{H}_D^{(3)} = \frac{\tilde{D}}{2} \sum_\mathbf{r} \sum_{\alpha,\beta=\pm 1} [\tilde{\mathcal{A}}_{0\alpha} \tilde{b}_{\mathbf{r}\beta}^\dagger \tilde{b}_{\mathbf{r}\beta} \tilde{b}_{\mathbf{r}\alpha} + h.c.], \tag{S7}$$

To simplify the notation, we will write a particular term of (39) (in momentum space) as $(I) = \tilde{b}_{\mathbf{q}_1,\alpha}^\dagger \tilde{b}_{\mathbf{q}_2,\beta} \tilde{b}_{\mathbf{q}_3,\gamma} f(\mathbf{q}_i, t)$, with

$$f(\mathbf{q}_{1,2,3}, t) = \begin{cases} 1 & t=0 \\ \gamma_{\mathbf{q}_3}^{xy} & t=1 \\ \gamma_{\mathbf{q}_3}^z & t=2 \end{cases}. \tag{S8}$$

The Nambu spinor of the bosonic operators can be Bogoliubov transformed into the quasi-particle representation $\vec{b}_\mathbf{k} = \mathcal{U}(\mathbf{k}) \vec{\beta}_\mathbf{k}$, where the matrix elements of $\mathcal{U}(\mathbf{k})$ are obtained from the Bogoliubov coefficients given in Eq. (20)

$$\mathcal{U}(\mathbf{k}) = \begin{pmatrix} \mathcal{U}_{2\times 2}^{11}(\mathbf{k}) & \mathcal{U}_{2\times 2}^{12}(\mathbf{k}) \\ \mathcal{U}_{2\times 2}^{21}(\mathbf{k}) & \mathcal{U}_{2\times 2}^{22}(\mathbf{k}) \end{pmatrix} \tag{S9}$$

$$= \begin{pmatrix} u_{k,+1} & 0 & v_{k,+1} & 0 \\ 0 & u_{k,-1} & 0 & v_{k,-1} \\ v_{k,+1} & 0 & u_{k,+1} & 0 \\ 0 & v_{k,-1} & 0 & u_{k,-1} \end{pmatrix}.$$

After applying the above-mentioned Bogoliubov transformation, we obtain

$$\begin{aligned}
(I) = \sum_{n_{1,2,3}} \{ & F^{(a)}(\alpha\beta\gamma, n_{1,2,3}; \mathbf{q}_{1,2,3}, t) \beta_{\mathbf{q}_1, n_1} \beta_{\mathbf{q}_2, n_2} \beta_{\mathbf{q}_3, n_3} \\
& + F^{(b)}(\alpha\beta\gamma, n_{1,2,3}; \mathbf{q}_{1,2,3}, t) \beta^\dagger_{\bar{\mathbf{q}}_1, n_1} \beta^\dagger_{\bar{\mathbf{q}}_2, n_2} \beta^\dagger_{\bar{\mathbf{q}}_3, n_3} \\
& + F^{(c)}(\alpha\beta\gamma, n_{1,2,3}; \mathbf{q}_{1,2,3}, t) \beta^\dagger_{\bar{\mathbf{q}}_1, n_1} \beta^\dagger_{\bar{\mathbf{q}}_2, n_2} \beta_{\mathbf{q}_3, n_3} \\
& + F^{(d)}(\alpha\beta\gamma, n_{1,2,3}; \mathbf{q}_{1,2,3}, t) \beta^\dagger_{\bar{\mathbf{q}}_1, n_1} \beta_{\mathbf{q}_2, n_2} \beta_{\mathbf{q}_3, n_3} \\
& + L^{(c)}(\alpha\beta\gamma, n_{1,2,3}; \mathbf{q}_{1,2,3}, t) \delta_{\mathbf{q}_3, \bar{\mathbf{q}}_2} \delta_{n_3, n_2} \beta^\dagger_{\bar{\mathbf{q}}_1, n_1} \\
& + L^{(d)}(\alpha\beta\gamma, n_{1,2,3}; \mathbf{q}_{1,2,3}, t) \delta_{\mathbf{q}_2, \bar{\mathbf{q}}_1} \delta_{n_2, n_1} \beta_{\mathbf{q}_3, n_3} \}.
\end{aligned} \quad (S10)$$

The explicit forms of $F^{(a,b,c,d)}$ and $L^{(c,d)}$ can be obtained by simple algebras, which are not shown here for the sake of brevity.

The "sink" ("source") function $F^{(a)}$ ($F^{(b)}$) is symmetric under permutations of all three legs (momenta and flavors). Consequently, we introduce the symmetrized functions

$$\tilde{F}^{(a)} \equiv \sum_{P(\mathbf{q}_{1,2,3}; n_{1,2,3})} F^{(a)}, \qquad \tilde{F}^{(b)} \equiv \sum_{P(\mathbf{q}_{1,2,3}; n_{1,2,3})} F^{(b)}. \quad (S11)$$

Similarly, the "decay" function $F^{(c)}$ and the "fusion" function $F^{(d)}$ are symmetrized for the two outgoing and the two incoming legs, respectively,

$$\tilde{F}^{(c)} = \sum_{P(\mathbf{q}_{1,2}; n_{1,2})} F^{(c)}, \qquad \tilde{F}^{(d)} = \sum_{P(\mathbf{q}_{2,3}; n_{2,3})} F^{(c)}. \quad (S12)$$

After inserting the above results into Eq. (39), we obtain the explicit forms of the cubic vertices in $V^{(3)}_{s/d}$ Eq. (24) and $V^L_\alpha$ in Eq. (25).

**Supplementary Note 7. Quartic vertex**

The quartic contributions to the expansion (9) are

$$\mathcal{H}^{(4)} = \mathcal{H}^{(4)}_{intra} + \mathcal{H}^{(4)}_{inter}, \quad (S13)$$

with

$$\begin{aligned}
\mathcal{H}^{(4)}_{intra} = \tilde{J} \sum_{\langle \mathbf{r}, \mathbf{r}' \rangle, \nu} \sum_{\alpha, \beta = \pm 1} \{ & [a_\nu \tilde{S}^\nu_{00} \tilde{S}^\nu_{00} \tilde{b}^\dagger_{\mathbf{r}\alpha} \tilde{b}^\dagger_{\mathbf{r}'\beta} \tilde{b}_{\mathbf{r}\alpha} \tilde{b}_{\mathbf{r}'\beta}] \\
& + \sum_{\alpha'\beta' = \pm 1} [a_\nu \tilde{S}^\nu_{\alpha\beta} \tilde{S}^\nu_{\alpha'\beta'} \tilde{b}^\dagger_{\mathbf{r}\alpha} \tilde{b}^\dagger_{\mathbf{r}'\alpha'} \tilde{b}_{\mathbf{r}\beta} \tilde{b}_{\mathbf{r}'\beta'}] \\
& - 2 \sum_{\alpha' = \pm 1} [a_\nu \tilde{S}^\nu_{\alpha\beta} \tilde{S}^\nu_{00} \tilde{b}^\dagger_{\mathbf{r}\alpha} \tilde{b}^\dagger_{\mathbf{r}'\alpha'} \tilde{b}_{\mathbf{r}\beta} \tilde{b}_{\mathbf{r}'\alpha'}]
\end{aligned} \quad (S14)$$

$$-\sum_{\alpha'=\pm 1}[a_\nu \tilde{S}^\nu_{\alpha 0}\tilde{S}^\nu_{\beta 0}\tilde{b}^\dagger_{\mathbf{r}\alpha}\tilde{b}^\dagger_{\mathbf{r}'\beta}\tilde{b}^\dagger_{\mathbf{r}'\alpha'}\tilde{b}_{\mathbf{r}'\alpha'} + h.c.]$$

$$-\sum_{\alpha'=\pm 1}[a_\nu \tilde{S}^\nu_{\alpha 0}\tilde{S}^\nu_{0\beta}\tilde{b}^\dagger_{\mathbf{r}\alpha}\tilde{b}^\dagger_{\mathbf{r}'\alpha'}\tilde{b}_{\mathbf{r}'\alpha'}\tilde{b}_{\mathbf{r}'\beta} + h.c.]\}.$$

Similarly to the cubic contribution, $\mathcal{H}^{(4)}_{inter}$ can be obtained from $\mathcal{H}^{(4)}_{intra}$ by substituting $\tilde{J} \to \tilde{J}'$, $a_\nu \to b_\nu$. The matrix elements appear in the normal ordering of the quartic vertex are defined as:

$$\begin{aligned}
\bar{N}^{\alpha\beta}_{\mathbf{rr}'} &\equiv \frac{1}{N}\sum_{\langle \mathbf{r},\mathbf{r}'\rangle} \langle b^\dagger_{\mathbf{r}\alpha} b_{\mathbf{r}'\beta}\rangle \\
&= \frac{1}{N}\sum_{\mathbf{k}}\sum_n \mathcal{U}^{21}_{\alpha,n}(\mathbf{k})[\mathcal{U}^{21}_{\beta,n}(\mathbf{k})]^* \cos[\mathbf{k}\cdot(\mathbf{r}'-\mathbf{r})], \\
\Delta^{\alpha\beta}_{\mathbf{rr}'} &\equiv \frac{1}{N}\sum_{\langle \mathbf{r},\mathbf{r}'\rangle} \langle b_{\mathbf{r}\alpha} b_{\mathbf{r}'\beta}\rangle \\
&= \frac{1}{N}\sum_{\mathbf{k}}\sum_n \mathcal{U}^{11}_{\alpha,n}(\mathbf{k})[\mathcal{U}^{21}_{\beta,n}(\mathbf{k})]^* \cos[\mathbf{k}\cdot(\mathbf{r}'-\mathbf{r})], \\
\bar{\Delta}^{\alpha\beta}_{\mathbf{rr}'} &\equiv \frac{1}{N}\sum_{\langle \mathbf{r},\mathbf{r}'\rangle} \langle b^\dagger_{\mathbf{r}\alpha} b^\dagger_{\mathbf{r}'\beta}\rangle \\
&= \frac{1}{N}\sum_{\mathbf{k}}\sum_n \mathcal{U}^{21}_{\alpha,n}(\mathbf{k})[\mathcal{U}^{11}_{\beta,n}(\mathbf{k})]^* \cos[\mathbf{k}\cdot(\mathbf{r}'-\mathbf{r})],
\end{aligned} \tag{S15}$$

We note that some of these matrix elements are equal to zero because of the residual U(1) symmetry of the antiferromagnetic order. To obtain the normal-ordered Hamiltonian Eq. (34), we apply a mean-field (Hartree-Fock) decoupling to the quartic Hamiltonian Eq. (42), for example,

$$\begin{aligned}
\tilde{b}^\dagger_{\mathbf{r}\alpha}\tilde{b}^\dagger_{\mathbf{r}'\beta}\tilde{b}_{\mathbf{r}\alpha}\tilde{b}_{\mathbf{r}'\beta} &\simeq \Delta^{\alpha\beta}_{\mathbf{rr}'}\tilde{b}^\dagger_{\mathbf{r}\alpha}\tilde{b}^\dagger_{\mathbf{r}'\beta} + \bar{N}^{\beta\beta}_{\mathbf{r}'\mathbf{r}'}\tilde{b}^\dagger_{\mathbf{r}\alpha}\tilde{b}_{\mathbf{r}\alpha} \\
&+ \bar{N}^{\beta\alpha}_{\mathbf{r}'\mathbf{r}}\tilde{b}^\dagger_{\mathbf{r}\alpha}\tilde{b}_{\mathbf{r}\beta} + \bar{\Delta}^{\alpha\beta}_{\mathbf{rr}'}\tilde{b}_{\mathbf{r}\alpha}\tilde{b}_{\mathbf{r}'\beta} \\
&+ \bar{N}^{\alpha\alpha}_{\mathbf{rr}}\tilde{b}^\dagger_{\mathbf{r}'\beta}b_{\mathbf{r}'\beta} + \bar{N}^{\alpha\beta}_{\mathbf{rr}'}\tilde{b}^\dagger_{\mathbf{r}'\beta}\tilde{b}_{\mathbf{r}\alpha}.
\end{aligned} \tag{S16}$$

The coefficients $V^{(4,N)}_{\alpha\alpha}$ that appear in the normal term of Eq. (34) can be derived after consecutive Fourier and Bogoliubov transformations.

**Supplementary Note 8. One-loop diagrams in the long-wavelength limit**

Without loss of generality, we consider an isotropic Heisenberg model, i.e. $\tilde{J} = \tilde{J}'$, $\tilde{\Delta} = \tilde{\Delta}'$ to show the $1/\mathbf{q}$ divergence of the one-loop diagrams involving the Goldstone mode. According to Eq. (20),

$$\lim_{\mathbf{q}\to 0} u_{\mathbf{q},+}, v_{\mathbf{q},+} = \sqrt{\frac{\tilde{J}d}{v_{0,+}}}\frac{1}{\sqrt{q}}, -\sqrt{\frac{\tilde{J}d}{v_{0,+}}}\frac{1}{\sqrt{q}} \tag{S17}$$

where $v_{0,+} = 2\tilde{J}d\sqrt{\tilde{D}/(4\tilde{J}d^2) + 1/d}$ is the spin wave velocity of the Goldstone mode and $d = 3$ is the spatial dimension of the lattice equal to half of the coordination number. Note that the cubic vertices are proportional to a product of the Bogoliubov coefficients of three legs

$$V_{d,s}^{(3)} \propto u(v)_{\mathbf{q}_1,\alpha} u(v)_{\mathbf{q}_2,\beta} u(v)_{\mathbf{q}_3,\gamma}. \tag{S18}$$

For the decay and sink diagrams shown on the second line of figure 5a in the main text, we can choose, for instance, $\mathbf{q}_3 = \mathbf{q} \sim \mathbf{0}, \gamma = +1$ to contract with the leg of the long-wavelength bosons. Consequently,

$$\Sigma^{(d,s)}(+) \sim |V_{d,s}^{(3)}(\mathbf{q}_{1,2}, \mathbf{q}; \alpha\beta+)|^2 \sim (1/\sqrt{\mathbf{q}})^2 \sim 1/q. \tag{S19}$$

As for the cubic-linear diagrams, we need to choose two legs to contract with the long-wavelength boson, implying that

$$\Sigma^{(cl)}(+) \sim V_d^{(3)}(\mathbf{0}, -\mathbf{q}\mathbf{q}; - + +) \sim 1/q. \tag{S20}$$

Finally, notice that $V_{++}^{(4,N)} \sim 1/q$ in the long-wavelength limit, because the quadratic forms of the transverse boson in Eq. (43) after the Bogoliubov transformation are proportional to $u(v)_{\mathbf{q},+1} u(v)_{\mathbf{q},+1}$. By adding up all diagrams in $\mathcal{O}(M^0)$, we have verified that the coefficient of the $1/q$-factor vanishes, implying that the Goldstone mode is preserved after the one-loop correction.

**Supplementary Note 9. Calculation of the inelastic neutron scattering intensity**

The imaginary-time dynamical spin susceptibility is defined as:

$$\chi^{\mu\nu}(\mathbf{q}, i\omega_n) = -\int_0^\beta d\tau e^{i\omega_n \tau} \langle \mathcal{T}_\tau [s_\mathbf{q}^\mu(\tau) s_{\bar{\mathbf{q}}}^\nu(0)] \rangle. \tag{S21}$$

The *real-time* spin-spin correlation function in Eq. (22) is obtained by using the fluctuation-dissipation theorem at $T = 0$ after the analytic continuation $i\omega_n \to \omega + i0^+$:

$$S^{\mu\nu}(\mathbf{q}, \omega) = -2\Im[\chi^{\mu\nu}(\mathbf{q}, \omega)]. \tag{S22}$$

Up to order $\mathcal{O}(1/M)$, $S^{\mu\nu}$ acquires two contributions:

$$S^{\mu\nu} = S_{\text{qp}}^{\mu\nu} + S_{\text{tc}}^{\mu\nu}, \tag{S23}$$

where $S_{\text{qp}}^{\mu\nu}$ includes contributions from the quasi-particle channel associated with "transverse fluctuations" of the SU(3) order parameter, while $S_{\text{tc}}^{\mu\nu}$ includes two-particle contributions associated with "longitudinal fluctuations" of the SU(3) order parameter[3]. The latter is not analyzed in this work, as it only contributes to the continuum. After the one-loop corrections, the quasi-particle channel can be written as a linear combination of the dressed bosonic propagators $\mathcal{G}$. For details, see Ref. 3.

To explain the experimental data, we used the Lorentzian broadening line-shape on the calculation, which is naturally implemented from the Green's functions incorporated by simply adding an imaginary part to the real frequency: $\omega \to \omega + i\eta$. Since the instrumental resolution of the INS spectrum is conventionally modeled with a Gaussian function, here we confirm the validity of using a Lorentzian broadening in our analysis. Supplementary Figure 5a compares the resolution convoluted Gaussian and Lorentzian broadening for the same calculated energy scan at the ZC ($\mathbf{Q}=\mathbf{Q_m}$). For clearer comparison, the $T_1$ transverse mode was subtracted in the calculations. The comparison shows a discrepancy in the tail of the $T_2$-mode, however, the $L$-mode has nearly identical linewidths and peak-areas for both broadenings, which confirms that the Lorentzian and Gaussian broadenings essentially give the same result. Also, this result shows that the slightly extended tail of $T_2$-mode does not affect the line shape of the $L$-mode at the ZC. Therefore, we conclude that the Lorentzian broadening provides a good approximation of the resolution function in the present case.

**Supplementary Note 10. Longitudinal mode extraction at the ZC**

Since the $L$-mode is close to the large quasi-elastic spectral weight in the proximity of the magnetic Bragg peak at the ZC ($\mathbf{Q}=\mathbf{Q_m}$), the modeling of this scattering may affect the extraction of the accurate peak position of the $L$-mode. We note that a gaped $T_1$-transverse mode can further complicate the determination of the quasi-elastic line-shape. A single Gaussian (Lorentzian) function underestimate (overestimate) the line width. Alternatively, we modeled the peak with the Gaussian + Lorentzian functions. Here the Gaussian peak describes the elastic scattering and the Lorentzian peak with its detailed balance describe the inelastic contribution. Supplementary Figure 5b shows the fitting the function to the data (cyan solid line). The longitudinal- (transverse-) mode was extracted from the remaining spectral weight, which gives the peak center at 1.30 meV with FWHM=0.644 meV (2.53 meV with FWHM=0.23).

**Supplementary Note 11. Fitting the INS spectrum.**

Since the one-loop corrections involve the evaluation of several numerical integrations (cf. Eqs. (25, 32, 33) and Eq. (S15)), and the many-body effects are stronger near the zone center, we adopted the criterion of reducing the number of free model parameters to a minimum value. In addition, the splitting of the transverse modes and the longitudinal mode is most striking near the ZC and the calculated spectrum has a strong dependence on $\tilde{J}$ and $\tilde{D}$, while it is much less sensitive to $\tilde{\Delta}$ and $\tilde{\Delta}'$. Moreover, as shown in the Supplementary Figure 4a and 4d, the spin excitations along the $L$-direction are almost flat, indicating that the inter-layer exchange constant is relatively small. We then choose $\tilde{J}$ and $\tilde{D}$ as free parameters while fixing $\tilde{\Delta} = \tilde{\Delta}' = 1/3$ (isotropic exchange interaction $\Delta = \Delta' = 1$ in the $S = 2$ model Eq. (1)) and $\tilde{J}' = 0.1\tilde{J}$. The parameter set $\mathcal{A}$ (GLSWT) is obtained by fitting the energy of the longitudinal- and transiverse modes at the ZC with the analytical expression given in Eq. (21) (see Note 10 for the experimental energy values). The fitting procedure becomes more challenging upon inclusion of the one loop corrections because the calculation involves multiple integrations and the renormalization in the Hamiltonian parameters turns out to be rather strong for $Ba_2FeSi_2O_7$. Here, we simply compute the renormalization of the real part of the self-energy for both modes at the ZC and then deform the set of parameters $\mathcal{A}$ (GLSWT) until the renormalized peaks positions match with the experimental peak positions at the ZC. The best-fit parameter set is listed as set $\mathcal{B}$ in Table 1. The errors of these parameters were estimated from the uncertainty in the $\chi^2$ value[4] defined by $\chi^2 = \frac{(E_L^{exp.}-E_L^{cal.})^2}{\sigma_{std.dev.}^2} + \frac{(E_{T2}^{exp.}-E_{T2}^{cal.})^2}{\sigma_{std.dev.}^2}$, where $E_i^{exp.}$, $E_i^{cal.}$, $\sigma_{std.dev.}$ ($i$=L, $T_2$) corrrespond to the energy and the standard deviation of the modes at the ZC.

**Supplementary Note 12. Ordered moment**

To have an independent validation of the model parameters obtained from fits of the INS data, we also compare the calculated staggered magnetic moment $M_S$ using the set $\mathcal{B}$ given in Table 1 with the value of 2.95 $\mu_B$ that was extracted from the neutron diffraction experiment[1]. At the mean-field level, the effective $S = 1$ model predicts $M_S = g_{ab} \sqrt{3} |\langle s_{\mathbf{r}}^x \rangle| \mu_B = 2.92\ \mu_B$ for $g_{ab} = 2.18$ [1], while the $1/M$ correction from the GLSWT yields $M_S = 2.79\ \mu_B$, where the factor $\sqrt{3}$ arises from $P_{S=1} S_{\mathbf{r}}^x P_{S=1} = \sqrt{3} s_{\mathbf{r}}^x$. We note that the relatively smaller calculated value of $M_S$ can be attributed to the fact that it is calculated from the effective $S = 1$ model. To verify this argument, we performed the mean field calculation using the original $S = 2$ model Eq. (1) and obtained $M_S = 3.09\ \mu_B$. We further infer that the $1/M$ correction, which corresponds to an SU(5) GLSWT calculation for $S = 2$, will bring the calculated value very close to the measured one (for a relative moment reduction equal to the one obtained for the SU(3) GLSWT the result is $M_S = 2.95\ \mu_B$ in good agreement with the measured value). This agreement confirms the validity of the spin one model obtained from fits of the INS data with the GLSWT plus one-loop corrections. The small reduction (-4.5%) of $M_S$, relative to the mean field value, indicates that the assumption of validity of a perturbative $1/M$-expansion is self-consistently verified for the Hamiltonian parameters of $Ba_2FeSi_2O_7$.

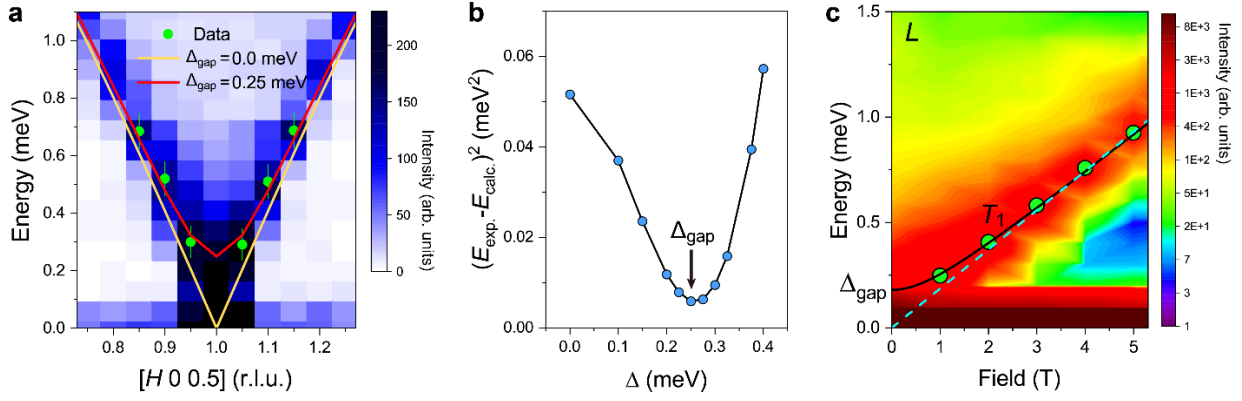

**Supplementary Figure 2. Gap excitation at magnetic zone center**

**a** Contour map for low energy inelastic neutron scattering data near the zone center. The magnon dispersion extracted from the data is indicated by green circles and is compared with the dispersion calculated with the GLSWT with and without a gap ($\Delta_{\text{gap}}$ =0.25 meV) in the spectrum. **b** Sum of the square of the energy difference between the measured and calculated dispersion as a function of gap size. The arrow marks the gap size $\Delta_{\text{gap}} \sim 0.25$ meV that best fits the data. **c** Contour plot for the field-dependent inelastic neutron scattering at ZC. The cyan circles indicate the energy of $T_1$ transverse modes. '$L$' denotes the scattering from the longitudinal mode. The $T_1$-modes were fitted to gapless linear function (dashed cyan line) and gapped parabola function (solid black line), $\omega_{mag} = \sqrt{\Delta_{gap}^2 + (c)^2 H^2}$. Fitting to the gapped parabola function gives a gap with 0.18 (2) meV.

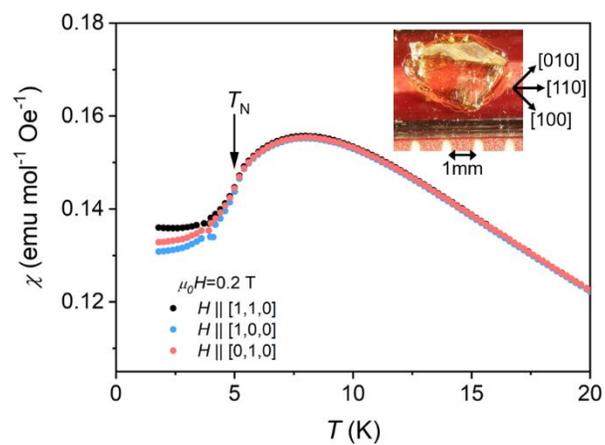

**Supplementary Figure 3. Magnetic susceptibility with angular field-dependence in *ab*-plane.**
Magnetic susceptibilities measured with magnetic fields ($H$=0.2 T) along $H$||[1, 0, 0], [1, 1, 0], and [0, 1, 0] directions of the crystal structure. The inset shows a picture of the measured $Ba_2FeSi_2O_7$ single crystal with the crystal orientations.

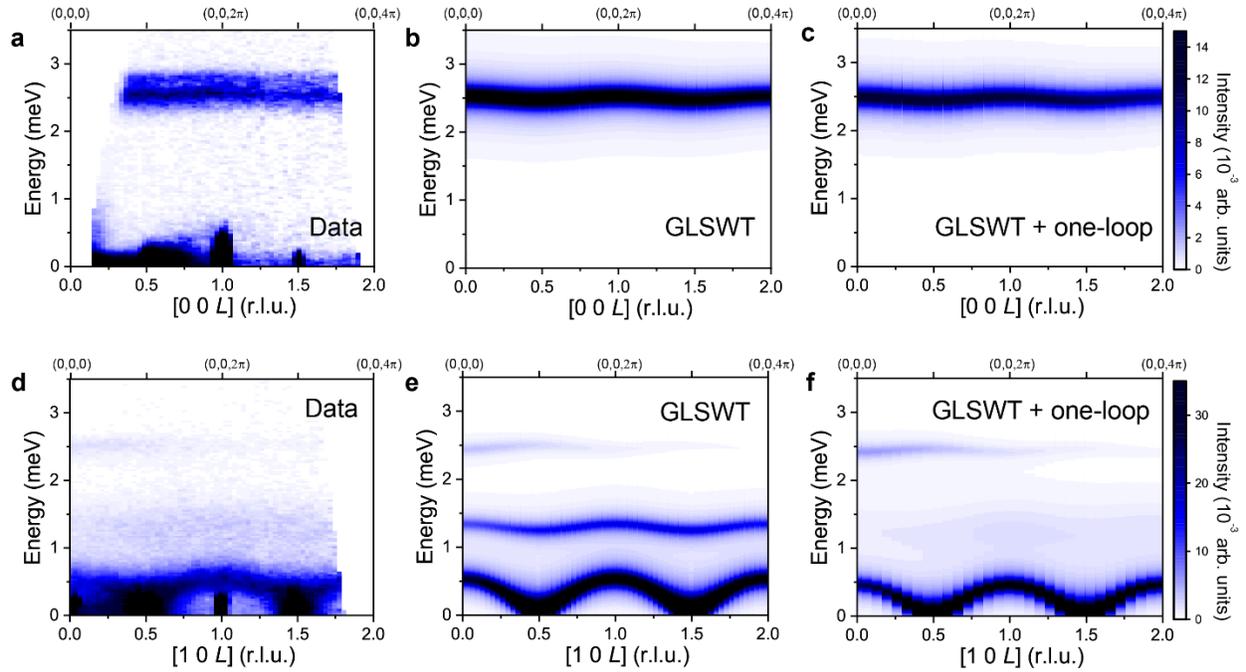

**Supplementary Figure 4. Spin excitation along *L*-direction**

Inelastic neutron scattering spectra along $[H, 0, L]$ for $H=0$ (**a-c**) and 1 (**d-f**) along with the calculated spectra using the GLSWT and GLSWT+one-loop corrections using parameter sets $\mathcal{A}$ and $\mathcal{B}$ in Table. 1, respectively. All the calculated spectra were convoluted with the instrumental resolution of HYSPEC.

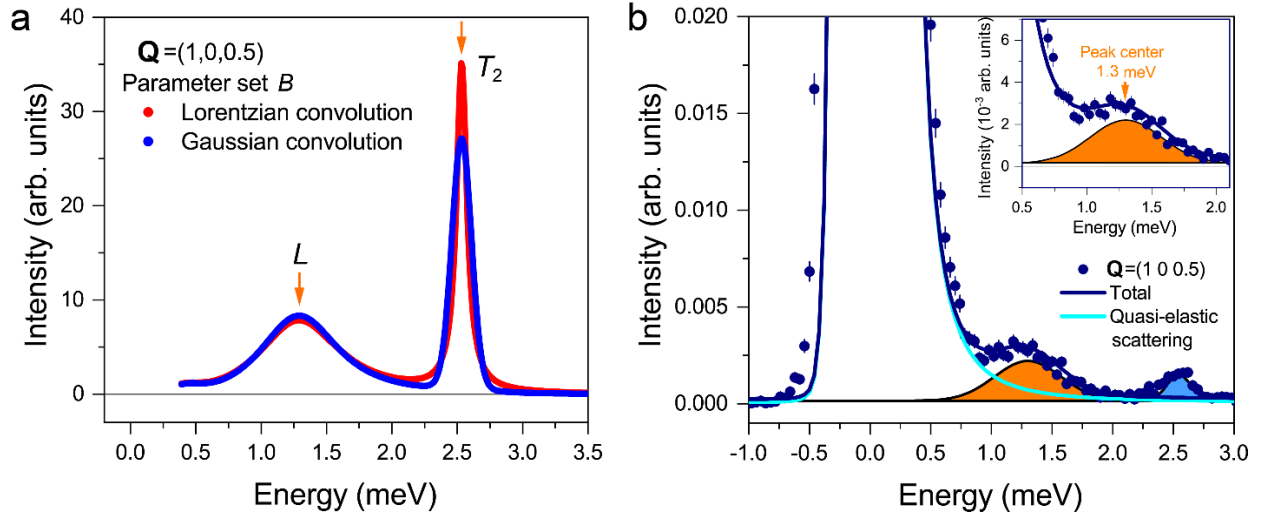

**Supplementary Figure 5. Line broadening of the calculated spectrum and the extraction of the longitudinal mode at the ZC.**
**a** Calculated INS intensity of GLSWT+one-loop correction model after applying the Lorentzian and Gaussian broadening (convolution). $L$ ($T_2$) indicates the longitudinal (transverse) mode. **b** Constant momentum cut at the ZC ($\mathbf{Q} = \mathbf{Q_m}$), measured using HYSPEC at SNS, shows a large elastic scattering, quasi-elastic scattering, $L$, and $T_2$ modes. The large elastic and quasi-elastic scattering are fitted with Gaussian + Lorentzian functions (solid cyan line). The orange- and blue-shaded regions indicate $L$- and $T_2$-modes extracted from fitting.

**References**
1. Jang T-H, *et al.* Physical properties of a quasi-two-dimensional square lattice antiferromagnet $Ba_2FeSi_2O_7$. Preprint at https://arxiv.org/abs/2108.00999 (2021).
2. Mai TT, *et al.* Terahertz spin-orbital excitations in the paramagnetic state of multiferroic $Sr_2FeSi_2O_7$ *Phys Rev B* **94**, (2016).
3. Mourigal M, Fuhrman WT, Chernyshev AL, Zhitomirsky ME. Dynamical structure factor of the triangular-lattice antiferromagnet. *Phys Rev B* **88**, 094407 (2013).
4. Bevington PR, Robinson DK. *Data reduction and error analysis*. McGraw Hill, New York (2003).